\documentclass[aps,twocolumn,nopacks,superscriptaddress,floatfix,prl,widetext]{revtex4-2}
\usepackage{graphicx}
\usepackage{color}
\usepackage{braket}
\usepackage{amsmath}
\usepackage{float}
\usepackage{subfigure}
\usepackage{placeins}
\usepackage{dcolumn}
\usepackage{bm}

\newcommand{\PRB}{Phys. Rev. B }

\begin{document}
\title{Polaron with Quadratic Electron-phonon Interaction}

\author{Stefano Ragni}
\affiliation{Faculty of Physics, Center for Computational Materials Science, University of Vienna,
A-1090 Vienna, Austria}
\author{Thomas Hahn}
\affiliation{Faculty of Physics, Center for Computational Materials Science, University of Vienna,
A-1090 Vienna, Austria}
\author{Zhongjin Zhang}
\affiliation{Department of Physics, University of Massachusetts, Amherst, Massachusetts 01003, USA}
\author{Nikolay Prokof’ev}
\affiliation{Department of Physics, University of Massachusetts, Amherst, Massachusetts 01003, USA}
\author{Anatoly Kuklov}
\affiliation{Department of Physics \& Astronomy, CSI, and the Graduate Center of CUNY, New York 10314, USA}
\author{Serghei Klimin}
\affiliation{TQC, Departement Fysica, Universiteit Antwerpen, Universiteitsplein 1, 2610 Antwerpen, Belgium}
\author{Matthew Houtput}
\affiliation{TQC, Departement Fysica, Universiteit Antwerpen, Universiteitsplein 1, 2610 Antwerpen, Belgium}
\author{Boris Svistunov}
\affiliation{Department of Physics, University of Massachusetts, Amherst, Massachusetts 01003, USA}
\affiliation{Wilczek Quantum Center, School of Physics and Astronomy and T. D. Lee Institute, Shanghai Jiao Tong University, Shanghai 200240, China}
\author{Jacques Tempere}
\affiliation{TQC, Departement Fysica, Universiteit Antwerpen, Universiteitsplein 1, 2610 Antwerpen, Belgium}
\author{Naoto Nagaosa}
\affiliation{RIKEN Center for Emergent Matter Science (CEMS),
2-1 Hirosawa, Wako, Saitama, 351-0198, Japan}
\affiliation{Department of Applied Physics, The University of Tokyo 7-3-1 Hongo, Bunkyo-ku,
Tokyo 113-8656, Japan}
\author{Cesare Franchini}
\affiliation{Faculty of Physics, Center for Computational Materials Science, University of Vienna,
A-1090 Vienna, Austria}
\affiliation{Dipartimento di Fisica e Astronomia, Universit\`a  di Bologna, 40127 Bologna, Italy}

\author{Andrey S. Mishchenko}
\affiliation{RIKEN Center for Emergent Matter Science (CEMS),
2-1 Hirosawa, Wako, Saitama, 351-0198, Japan}

\begin{abstract}
We present the first numerically exact study of a polaron with quadratic coupling to the oscillator displacement, using two alternative methodological developments. 
Our results cover both anti-adiabatic and adiabatic regimes and the entire range of electron-phonon coupling $g_2$, from the  system's stability threshold at attractive $g_2=-1$ to arbitrary strong repulsion at $g_2 \gg 1$. 
Key properties of quadratic polarons prove dramatically different from their linear counterparts. 
They (i) are insensitive even to large quadratic coupling except in the anti-adiabatic limit near the threshold of instability at attraction; (ii) depend only on the adiabatic ratio but are insensitive to the electron dispersion and dimension of space; (iii) feature weak lattice deformations even at the instability point.     
Our results are of direct relevance to properties of electrons at low densities in polar materials, including recent proposals for their superconducting states.
\end{abstract}

\maketitle


The first results on polarons with quadratic coupling to phonons were reported in Refs.~\cite{Kuklov1989,Gogolin1991}, which explored properties of large-radius solitons in the adiabatic limit at strong coupling.
Indications that nonlinear coupling to atomic displacements is important were found in several materials such as doped manganites \cite{Manganites2017}, halide perovskites \cite{Perovski2021}, and quantum paraelectrics \cite{Kumar2021}.
Most notable is the unusual $T^2$ dependence of resistivity at high temperature, which was explained by considering electron-phonon interactions (EPI) with quadratic dependence on the phonon coordinates \cite{Kumar2021,nazaryan}. 
The soft vibrational modes in these materials are transverse optical (TO) phonons for which the linear EPI is suppressed in the long-wave limit. 
However, local electron density changes the potential acting on nearby atoms and this change may increase or decrease the local spring constants. 
Early suggestions that bi-phonon exchange could be an important pairing mechanism
at low doping \cite{Ngai} were recently revisited by quantifying and employing them for explaining the superconducting properties of SrTiO$_3$ \cite{Marel,STO_chandra,STO_kiseliov}.
While the treatment of the problem was perturbative, the dimensionless coupling constant was estimated to be of order unity, raising the question of consistency.

In the low-density limit---when the polaron physics is most relevant \cite{FermiBlockade}---the key assumption on which the Migdal-Eliashberg theory is based (irrelevance of vertex corrections at strong coupling), namely $E_F \gg \Omega$, where $E_F$ is the Fermi energy and $\Omega$ is the characteristic phonon frequency, fails. 
Thus, any quantitative study of strong EPI effects in this limit should start from precise calculations of basic polaron properties such as its energy, $E$, effective mass, $m_*$, and the quasiparticle residue, $Z$. 
We are aware of only few theoretical attempts to account for quadratic EPI beyond  perturbation theory. 
The original work  \cite{Kuklov1989,Gogolin1991} was based on a variational approach for large-radius soliton-type solutions.  
A nonperturbative momentum average approximation \cite{BerciuPRL2006} was used to study the interplay between linear (Holstein model \cite{Holstein}) and non-linear EPI at zero temperature in Refs.~\cite{BerciuEPL2013,BerciuPRB2013}.
Effects of non-linear EPI on the formation of charge density waves, superconductivity, and quasi-particle properties were investigated in a series of papers \cite{JohnstonEPL2015,JohnstonPRB2015,JohnstonComPhys2020}. 
These determinant Monte Carlo \cite{DeverPRB2013} studies considered finite clusters (up to $N=8\times8$ sites) in two dimensions at high electron density and finite temperature.
More recently, the interplay between linear and quadratic EPI in the Fr\"ohlich model of continuous space polarons was studied at zero temperature in Ref.~\cite{HoutputPRB2021} using variational Feynman's path integral method \cite{FeynmanVariat1955}.

All studies find that quadratic interaction with positive/negative coupling constant
decreases/increases the effective strength of the linear EPI. However, none of the previous work
was able to treat effects of strong quadratic coupling in the thermodynamic limit without approximations, or was investigating polaron properties for purely quadratic interaction.
Meanwhile, as was mentioned above, there exist important cases when coupling to soft transverse phonons has no linear terms in the long-wave limit, e.g. quantum paraelectrics \cite{Kumar2021} and optically pumped systems \cite{Millis2017,Millis2021}.

In this Letter, we employ two complementary numerically exact methods to solve the polaron model with quadratic coupling to atomic displacements, or $X^2$-polarons, at zero temperature. 
The first one is based on 
Feynman diagrams and is best suited for studying dispersive phonons in the regimes of weak and intermediate coupling. 
The second method---performing best at strong coupling and becoming particularly simple in the dispersionless regime ---works with the path-integral representation for both the electron and atomic displacements.
The two methods are new methodological advances that go well beyond previous developments. 
We explore both adiabatic and non-adiabatic limits and find that $X^2$-polarons remain well-defined all the way to the instability threshold and possess the remarkable ability (especially in the adiabatic case) to resist renormalization even in the extreme strong coupling limit.


\underline{\textit{Model}}.
The key difference between our Hamiltonian and the well-studied Holstein model \cite{Holstein} is the quadratic, instead of linear, coupling to the local oscillator coordinates, $X_i=[b_i^{\dagger} + b_i^{\,}]/ \sqrt{2M\Omega} \equiv 
x_i/\sqrt{2M\Omega}$, where 
$M$ and $\Omega$ are the oscillator mass and frequency, respectively
(we use standard notation for on-site creation/annihilation operators for 
harmonic modes and electrons):
\begin{equation}
H=-t\sum_{<ij>} a_{j}^{\dagger} a_{i}^{\,}
+ \Omega \sum_i b_i^{\dagger} b_i^{\,}
+\frac{\Omega}{4} g_2 \sum_{i} n_{i} [ b_i^{\dagger} + b_i^{\,}]^2 \, .
\label{H}
\end{equation}
Here $n_i=a_{i}^{\dagger} a_{i}^{\,}$ is the electron occupation number.
The first two terms describe the electron hopping between nearest neighbor sites on the simple cubic lattice (in what follows we take $t$ as the unit of energy) and the local vibration modes, respectively. 
We count oscillator energies from their ground states, and use the dimensionless constant $g_2$ to parameterize the coupling. 
By writing the local potential energy for $n_i=1$ as $M\Omega^2 [1+g_2] X_i^2/2$, we observe that (i) the model becomes unstable at $g_2\le -1$, implying that the radius of convergence for a perturbative treatment in powers of $g_2$ is unity, (ii) the oscillator 
frequency is renormalized to  
\begin{equation}
\tilde{\Omega} = r \Omega, \qquad  r=\sqrt{1+g_2}\,,
\label{Or}
\end{equation}
and (iii) its ground state energy shifts to $\Omega (r-1)/2$.

\underline{\textit{Momentum-space representation}} (see also Supplemental material \cite{Suppl}). The first scheme is based on the Diagrammatic Monte Carlo (DiagMC) technique introduced in Ref.~\cite{PS98} and further developed in Ref.~\cite{MPSS}. 
The imaginary-time Green's function,  $G(\mathbf{k}, \tau ) $, for momentum state $\mathbf{k}$ is sampled stochastically from the series expansion in powers of $g_2$ expressed as Feynman diagrams in terms of bare electron and phonon propagators. 
The difference between linear and quadratic coupling is a more complex set of diagram topologies consisting of a set of $n$-phonon loops because now each interaction vertex involves two phonons (instead of one) being emitted or absorbed (see Eq.~(\ref{H})). 

\begin{figure}[h]
    \begin{center}
        \includegraphics[width=8.5cm]{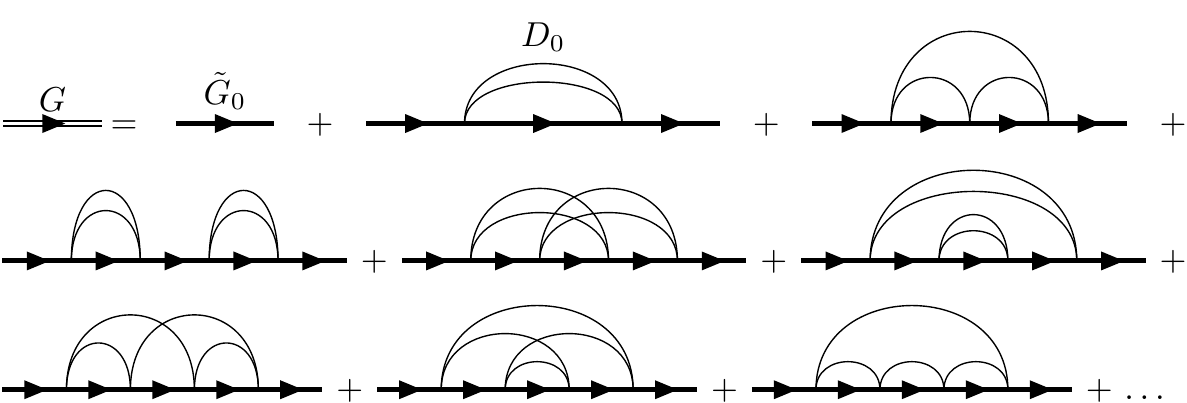}
    \end{center}
    \caption{\label{fig:topologies} Momentum-space diagrams for the Green's function up to fourth order in
   the quadratic coupling.}
\end{figure}

\begin{figure}[h]
    \begin{center}
        \includegraphics[width=8.0cm]{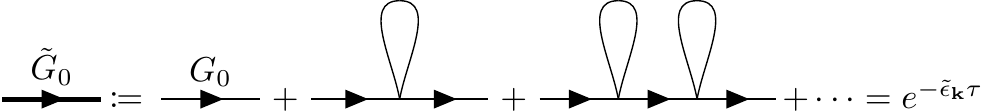}
    \end{center}
    \caption{\label{fig:loopsprop} 
    The geometric series of the 1-loop diagrams defines the ``Hartree" renormalized electron propagator. }
\end{figure}

Figure~\ref{fig:topologies} shows typical low-order diagrams. The simplest self-energy diagram is given by the 1-loop; its series, see Fig.~\ref{fig:loopsprop}, is absorbed into the ``bare" electronic propagator, $G_0 \to \tilde{G}_0$, by shifting the tight-binding dispersion $\epsilon_{\mathbf k} \to \tilde{\epsilon}_{\mathbf k} = \epsilon_{\mathbf k} +g_2 \Omega /4$:
\begin{equation}
\tilde{G}_0 (\mathbf{k}, \tau ) = e^{-\tilde{\epsilon}_{\mathbf k} \tau }, \qquad D_0 (\mathbf{q}, \tau ) = e^{-\Omega \tau } \,.
\label{Gr}
\end{equation}
The remaining high-order loops display a wide variety of topologies that increase dramatically with the order of the diagram. In addition, there exist multiple Wick 
pairings that are topologically equivalent and result in the same contribution, 
meaning that each diagram comes with the combinatorial factor $2^{N_\text{V}-N_2}$, where $N_\text{V}$ is the total number of vertices and $N_2$ is the number of 2-phonon loops, i.e. loops  consisting of two phonon propagators. 
The sign of the diagram is given by $(-g_2)^{N_\text{V}}$, meaning that the expansion 
is sign-positive for $g_2 < 0$.  

An ergodic sampling scheme includes the following updates: \\
\emph{Add/remove 2-loop}: Seed time, $\tau_1$ and $\tau_2$, and momentum, $\mathbf{q}_1$ and $\mathbf{q}_2$, variables for new vertices to be added to the diagram and balance this proposal by suggesting to remove any of the existing 2-loops. \\
\emph{Add/remove 3-loop}: This update is a straightforward generalization of the previous one, but for 3-loops. 
It is required for the generation of odd-order diagrams. \\
\emph{Relink}: Pick any two phonon propagators across the whole diagram at random such that they do not share vertices, i.e. they start and end on four different vertices. Propose a new diagram topology by connecting the four vertices with the two phonon propagators randomly. \\
Additional updates, such as changing time and/or momentum variables of the diagram, are introduced to improve the autocorrelation time.

Following Ref.~\cite{MPSS}, the simulation is extended to the $N$-phonon Green's function, which allows one 
to collect information about the structure of the phonon cloud. We employ standard procedures to extract the 
ground state energy, $E$, quasiparticle weight, $Z$, average number of phonons, $\langle N_{\rm ph} \rangle$, 
and effective mass, $m_*$, of the polaron.
%
\begin{figure*}
\subfigure{\includegraphics[width=0.32\textwidth]{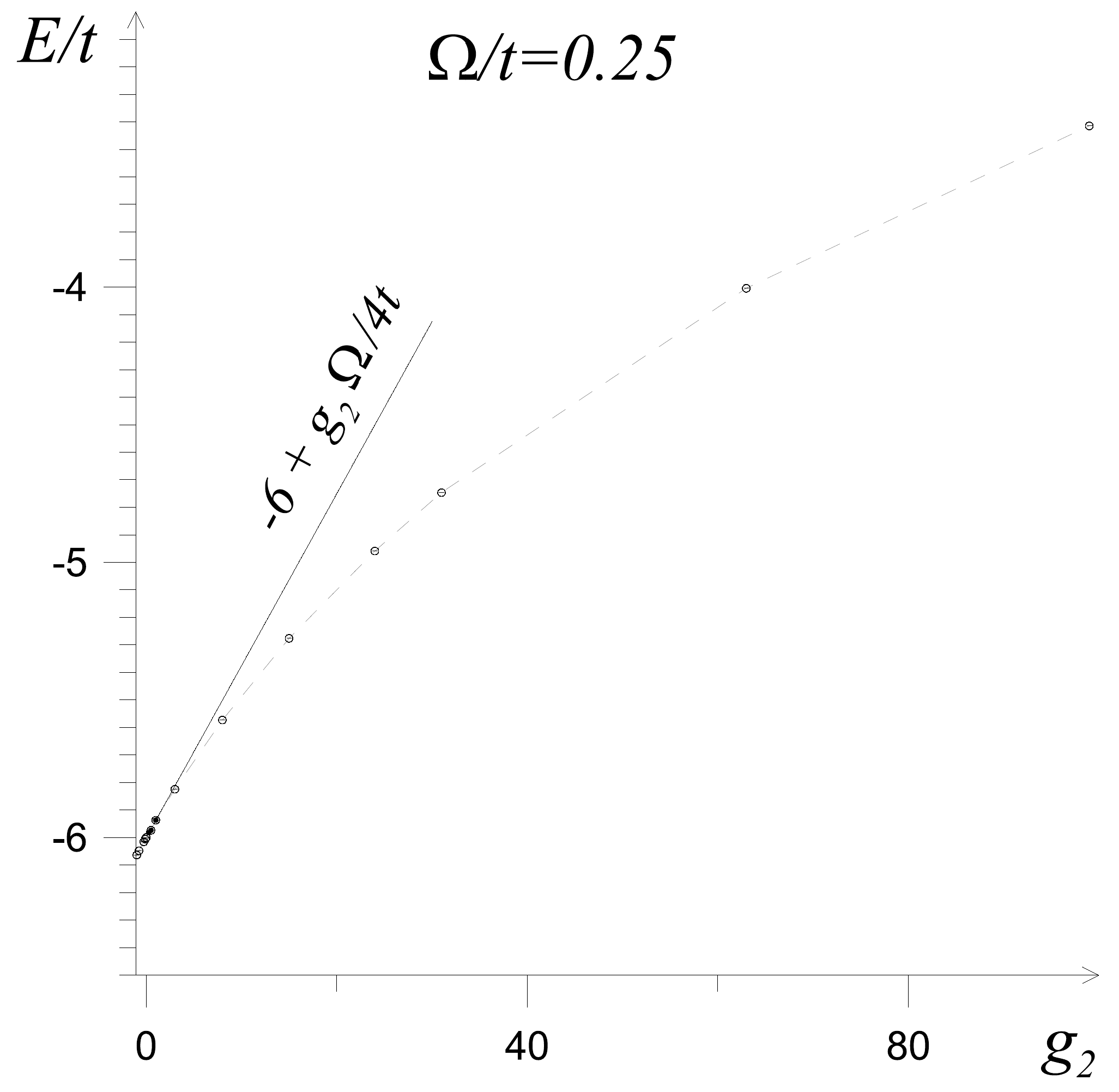}}
\subfigure{\includegraphics[width=0.32\textwidth]{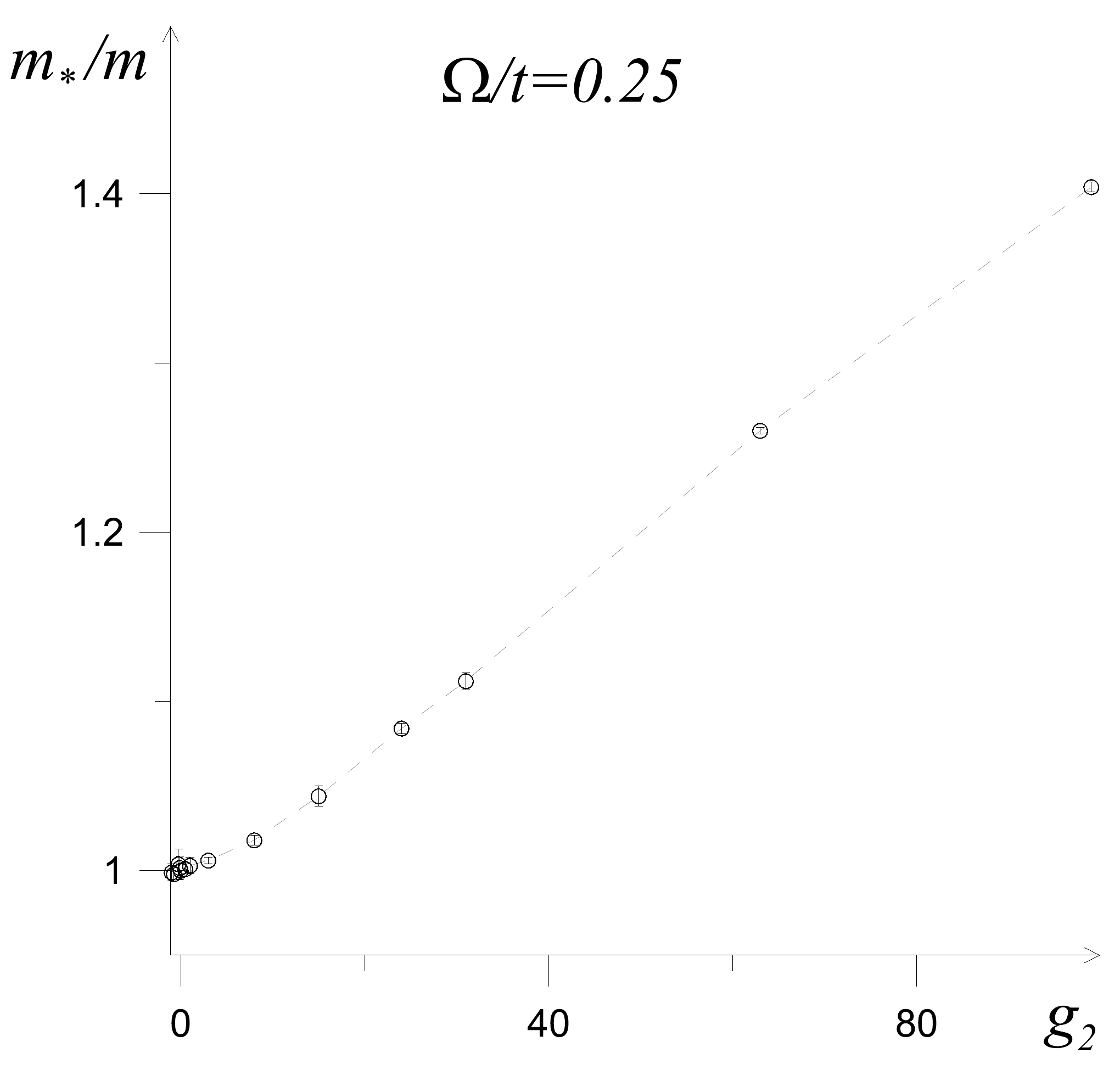}}
\subfigure{\includegraphics[width=0.32\textwidth]{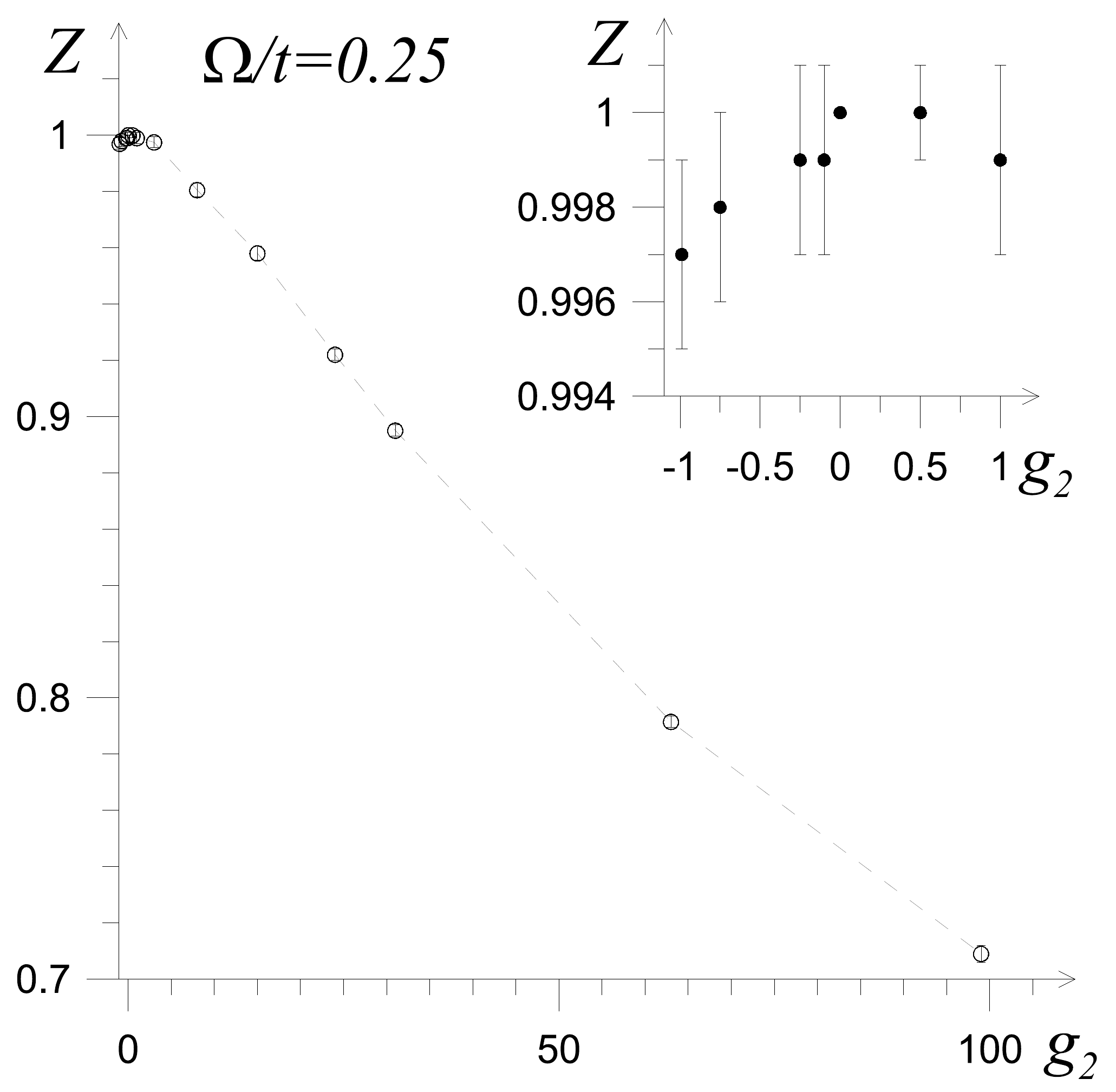}}
 \caption{\label{fig:Ad_EMZ} Polaron properties (energy, effective mass, and $Z$-factor)
in the adiabatic regime $\Omega/W=1/48$ (bandwidth $W=12t$ in 3D) as functions of EPI coupling.}
\end{figure*}


\underline{\textit{X-representation approach}}.
As discussed in Ref.~\cite{xrepresentation}, see also the Supplemental Material \cite{Suppl}, any non-linear coupling to local
vibration modes can be dealt with exactly and efficiently by specifying 
dimensionless atomic coordinates $x_i$ for all sites connected by the electron 
hopping transitions. The only new ingredient required for adapting the scheme of 
Ref.~\cite{xrepresentation} to the model in Eq.~(\ref{H}) is the imaginary-time 
oscillator propagator in the presence of the electron 
\begin{eqnarray}
\tilde{U}(y,x,\tau ) =
e^{(1/2) \Omega \tau  - \tilde{Q}(y,x,\tau) } \sqrt{ \frac{r} {4 \pi \sinh ( \tilde{\Omega} \tau)}  } \, \, , \nonumber\\
 \tilde{Q}(y,x,\tau)\, =\,
\frac{r [ \cosh ( \tilde{\Omega} \tau) (x^2+y^2) -2xy ]}{4 \sinh ( \tilde{\Omega} \tau ) } \, .
\label{Utilde}
\end{eqnarray}
For empty sites, the ``bare" propagator $U(y,x,\tau )$ has the same functional form
as $\tilde{U}(y,x,\tau )$ but with $r=1$ and $\tilde{\Omega} =\Omega$. Thus, for any electron's lattice path
and any set of atomic displacements on sites connected by hopping transitions, one has an 
exact sign-positive expression for the system's evolution operator subject to stochastic sampling without a bias.

\begin{figure*}[th]
\subfigure{\includegraphics[width=0.32\textwidth]{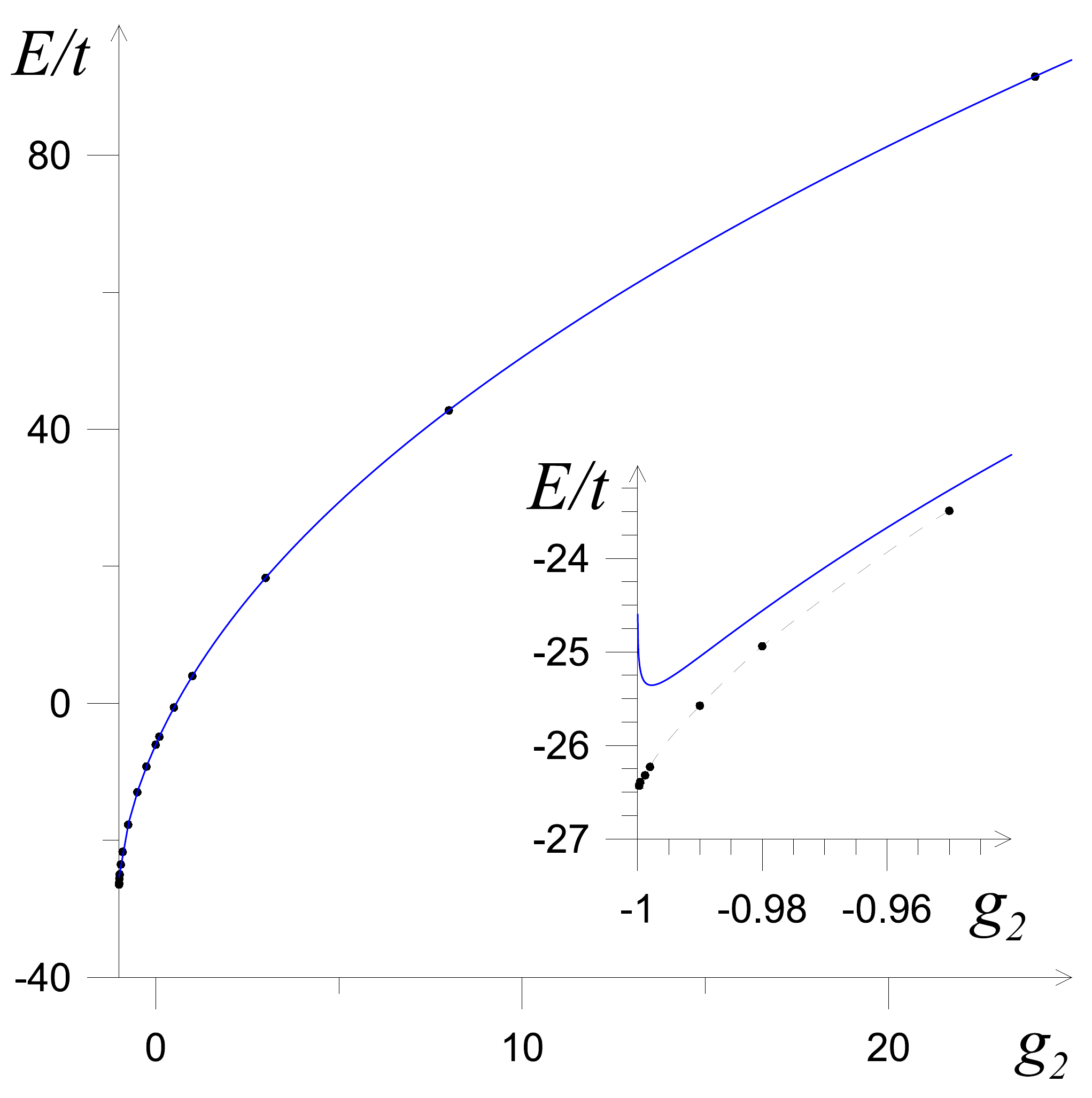}}
\subfigure{\includegraphics[width=0.32\textwidth]{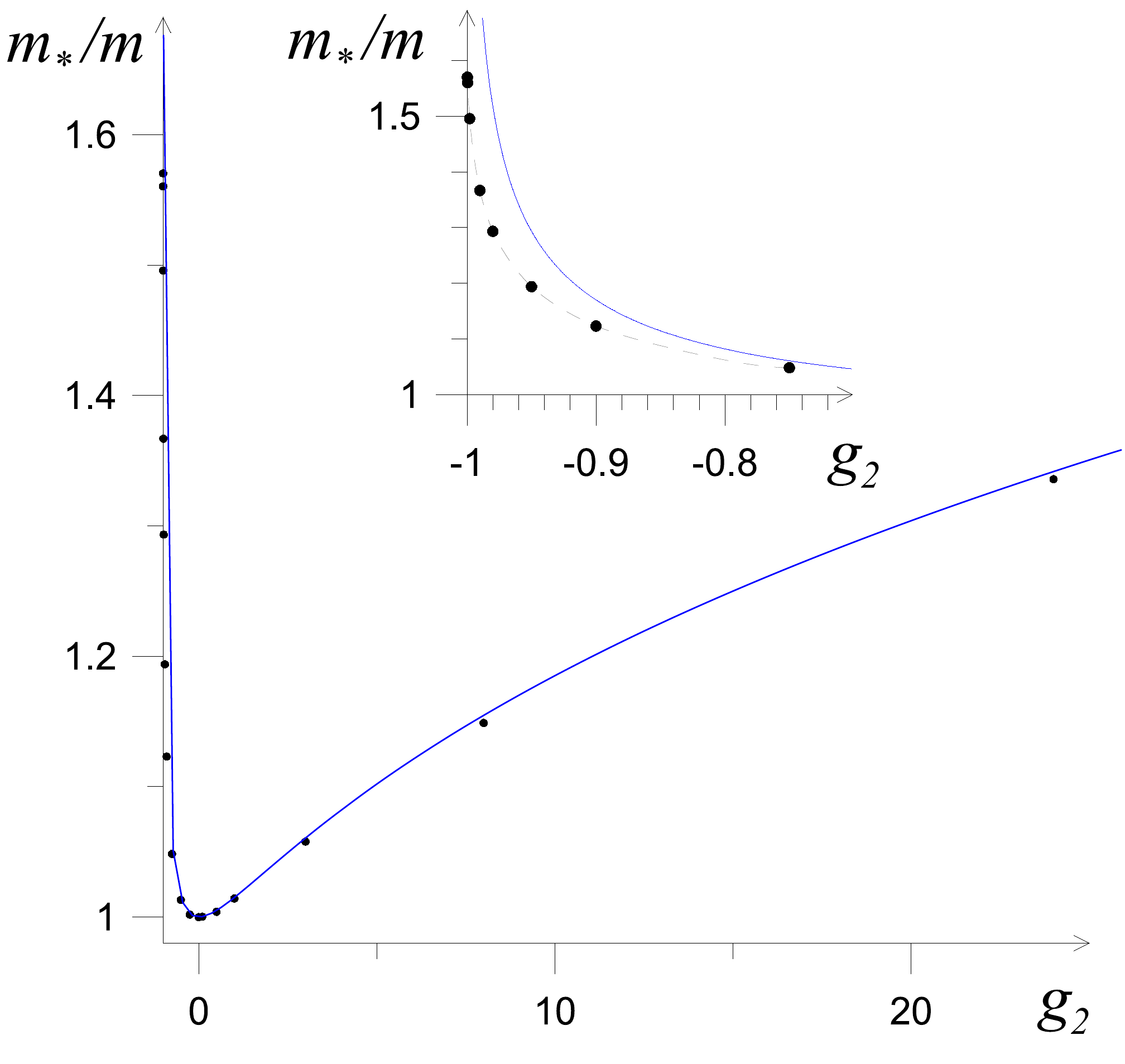}}
\subfigure{\includegraphics[width=0.32\textwidth]{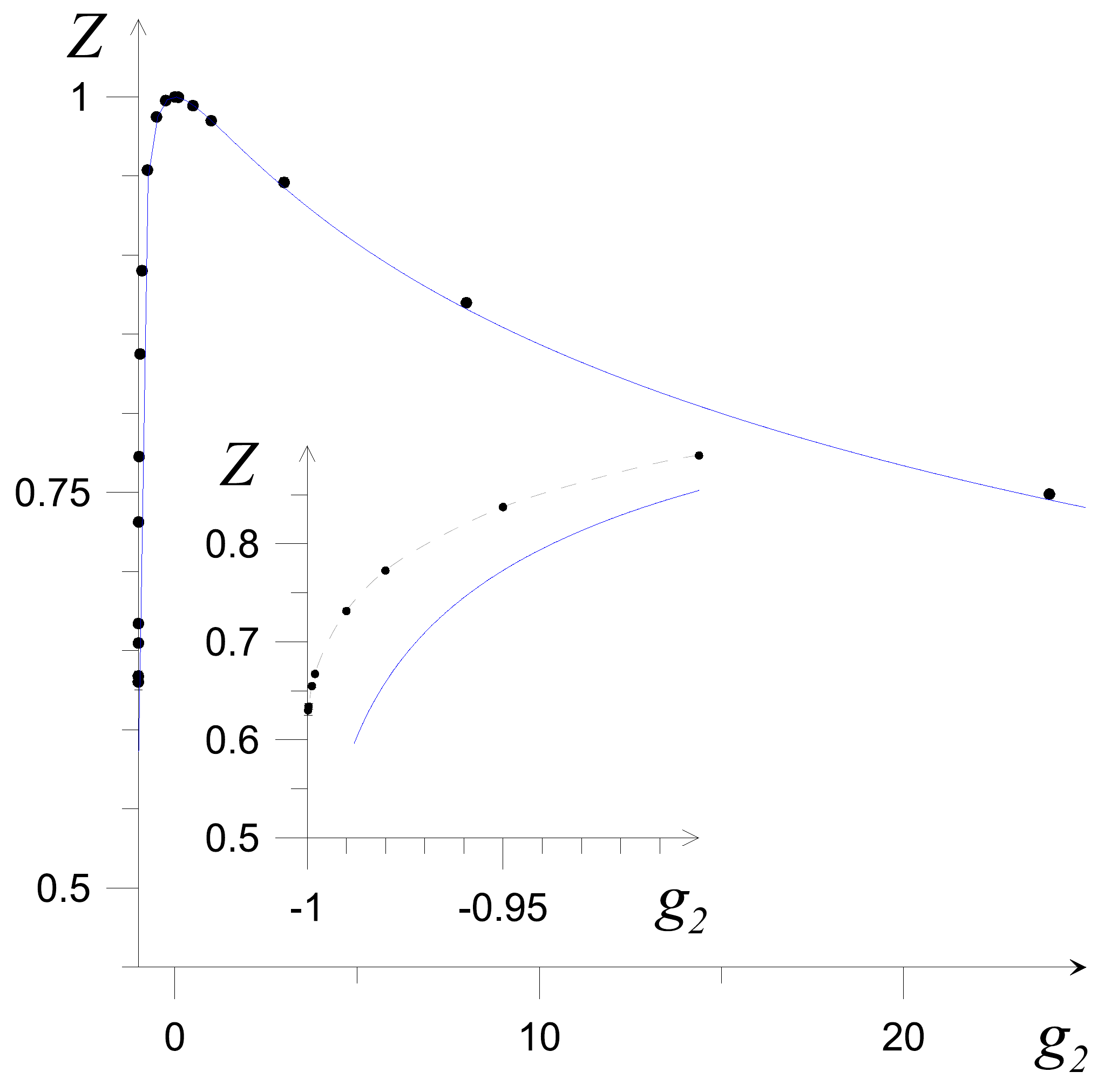}}
 \caption{\label{fig:Na_EMZ} Polaron properties (energy, effective mass, and $Z$-factor)
in the anti-adiabatic regime $\Omega/W=4$ in 3D as functions of EPI coupling.
Solid lines are the analytic predictions based on the atomic limit (\ref{ZEAlimit})--(\ref{Alimit}); dashed lines connecting data points
are used to guide an eye. }
\end{figure*}

In the  so-called atomic limit (AL),  $t=0$, the solution for the Green's 
function immediately follows from the Gaussian integral
$G_{A}(\tau ) = \int dx dy U(x) \tilde{U}(x,y,\tau) U(y)$ leading to
\begin{equation}
G_{A}(\tau )=Z_A e^{-E_A\tau} \left[ 1 - \left( \frac{1-r}{1+r}\right)^2 e^{-2\tilde{\Omega}\tau } \right]^{-1/2} \,,
\label{GAlimit}
\end{equation}
where 
\begin{equation}
Z_A = \frac{2\sqrt{r}}{1+r}\,, \qquad E_A=\frac{\Omega}{2}(r-1) \,.
\label{ZEAlimit}
\end{equation}
The spectral density, defined by  $G_A(\tau)=\int_0^{\infty} d\omega A_A(\omega ) e^{-\omega \tau}$, is readily obtained by Taylor expanding $G_A (\tau )$ in powers of 
$e^{-2\tilde{\Omega} \tau}$:
\begin{equation}
A_A(\omega ) = \sum_{k=0}^{\infty} Z_{2k} \delta (\omega - E_{2k}) \;.
\label{AAlimit}
\end{equation} 
It is a set of $\delta$-functions at frequencies $E_{2k} = E_A + 2k\tilde{\Omega} $
with $Z_{2k}$-factors equal to
\begin{equation}
Z_{2k} =Z_A \frac{(2k-1)!!}{2k!!}  \left[ \frac{1-r}{1+r} \right]^{2k} \; .
\label{Zkk}
\end{equation}
Finally, the atomic limit admits an exact solution for the average number of 
phonons in the polaron cloud defined as in Ref.~\cite{MPSS}
\begin{equation}
\langle N_{\rm ph} \rangle = \sum_{k=0}^{\infty} (2k) Z_{2k} = \frac{(1-r)^2}{4r}\;.
\label{NAlimit}
\end{equation} 

If both $\tilde{\Omega}$ and $\Omega$ are much larger than $t$ (meaning that to the leading approximation the oscillators always remain in the ground state as the electron moves),
then the finite-$t$ effects can be fully characterized by the overlap 
integral squared between the bare and renormalized ground states equal to $Z_A$:
\begin{equation}
E = - 6 Z_A t + E_A \,, \qquad m_*/m = Z_A^{-1} \;.
\label{Alimit}
\end{equation}

\begin{figure}[htb]
    \begin{center}
        \includegraphics[width=8.8cm]{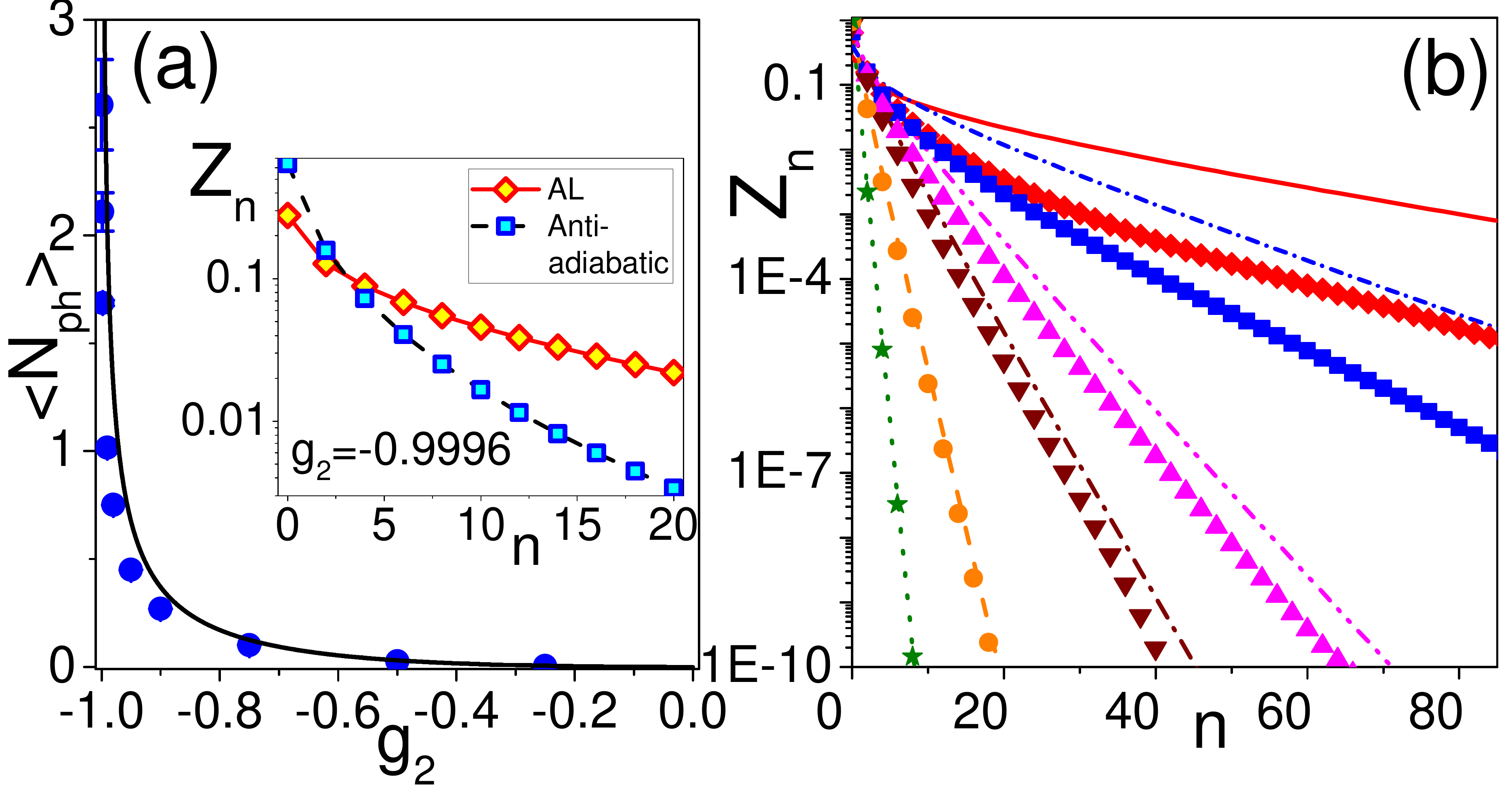}
    \end{center}
    \caption{\label{fig:fig5} 
    Phonon cloud in the anti-adiabatic limit in 3D at $\Omega/W=4$. 
    (a) Average number of phonons $\langle N_{\rm\scriptsize ph} \rangle$ (blue circles) and the AL prediction (line), Eq.~(\ref{NAlimit}), for negative coupling constant.
    (b) Phonon distributions $Z_n$ in the polaron cloud (symbols) and the AL prediction (lines, Eq.~(\ref{Zkk})), 
    $g_2=$: $-0.25$ (stars, dotted line), $-0.75$ (circles, dashed line), $-0.95$ (triangles down, dash-dot line), $-0.98$ (triangles up, dash-dot-dot line), $-0.998$ (squares, short-dash-dot line), and $-0.9996$ (diamonds, solid line). 
    Inset in panel (a) shows the phonon distribution (squares) and Eq.~(\ref{Zkk}) (diamonds) for $g_2=-0.9996$ 
    on a smaller scale. 
    }
\end{figure}

\underline{\textit{Results.}} 
In sharp contrast to conventional polarons, properties of $X^2$-polarons strongly depend on the sign of $g_2$, see Figs.~\ref{fig:Ad_EMZ} and \ref{fig:Na_EMZ}.
The effective mass  (quasiparticle weight) goes through a minimum (maximum) at $g_2 =0$, while the energy is an increasing function of $g_2$.
The asymmetry is especially notable in the anti-adiabatic case  $\Omega/W = 4$, see Fig.~\ref{fig:Na_EMZ}, and is directly linked to the fact that at $g_2 \to -1$ the local phonon frequency undergoes a dramatic change and ultimately softens to zero. Correspondingly, as long as the condition $\tilde{\Omega} >W$ is satisfied, the atomic-limit expressions featuring square-root singularities, see Eqs.~(\ref{GAlimit})--(\ref{Alimit}), provide an accurate description of the polaron properties. 
However, on approach to the stability threshold, this condition ultimately gets violated and the singularity is removed because an electron moves away before the slow phonon mode has a chance to adjust to its interacting ground state. 
This explains the remarkable fact, that all polaron properties remain well defined and regular in the $g_2 \to -1$ limit for any finite value of $t$, see Figs.~\ref{fig:Na_EMZ} and \ref{fig:fig5}. 

For positive $g_2$, all properties change gradually even at $g_2 \gg 1$. Moreover, in the adiabatic limit $\Omega/W = 1/48$, see Fig.~\ref{fig:Ad_EMZ}, both $Z$ and $m_*/m$ remain close to unity with sub percent accuracy for any $|g_2|\le 1$.   
One way to interpret the data is to make a  connection with the linear problem where the crossover from weak to strong coupling takes place when the dimensionless coupling $\lambda$, defined as the ratio between the coupling strength squared and the product of $W/2$ and $\Omega$, is of the order of unity. 
Introducing a similar parameter for model (\ref{H}) we get $\lambda  = 2(g_2 \Omega / 4)^2 /(W \Omega) = (g_2^2 \Omega ) / (8 W)$. 
Its value for $|g_2|=1$ is $\lambda  = 1/384$ for the adiabatic case shown in Fig.~\ref{fig:Ad_EMZ} and $\lambda  = 1/2$ for the anti-adiabatic case shown in Fig.~\ref{fig:Na_EMZ}.

Another stark difference between linear and quadratic couplings is found in the structure of the lattice distortion dragged along by polarons. 
It is quantified through probabilities $Z_n$ of having $n$ virtual phonons in the ground state [for AL it is given by Eq.~(\ref{Zkk})].
In the linear case,  the peak in $Z_n$ shifts from $n=0$ at weak coupling to large finite values of $n$ at strong coupling \cite{MPSS,Acoustic}. 
In contrast, $Z_n$ for $X^2$-polarons is peaked at $n=0$ and decreases exponentially at large  $n$ (Fig.~\ref{fig:fig5}b) for any value of negative $g_2$, including the close vicinity of the instability point $(1+g_2) < 10^{-3}$ when the average phonon number of phonons $\langle N_{\rm ph} \rangle$ is already large (Fig.~\ref{fig:fig5}a).  
Somewhat counter-intuitively, $Z_{n=0}$ for the moving particle ($t>0$) is larger than for the localized particle in the AL ($t=0$), whereas for large $n$ the opposite is true, see inset in  Fig.~\ref{fig:fig5}a. 
Only for large $g_2$ in the deep adiabatic limit $\Omega /t \ll 1$ it is possible that $Z_n$ has a peak at finite $n$ due to formation of the soliton state \cite{Kuklov1989}.

We also find that properties of quadratic polarons mostly depend on the particle bandwidth, and are rather insensitive to the form of the dispersion relation and even the dimension of space,  e.g., $E$, $Z$, $\langle N_{\rm ph} \rangle$, and $m_*$ are practically indistinguishable between the 3D and 1D cases provided the bandwidth is the same 
(see Supplemental material \cite{Suppl}).            

\begin{figure}[htb]
    \begin{center}
        \includegraphics[width=8.cm]{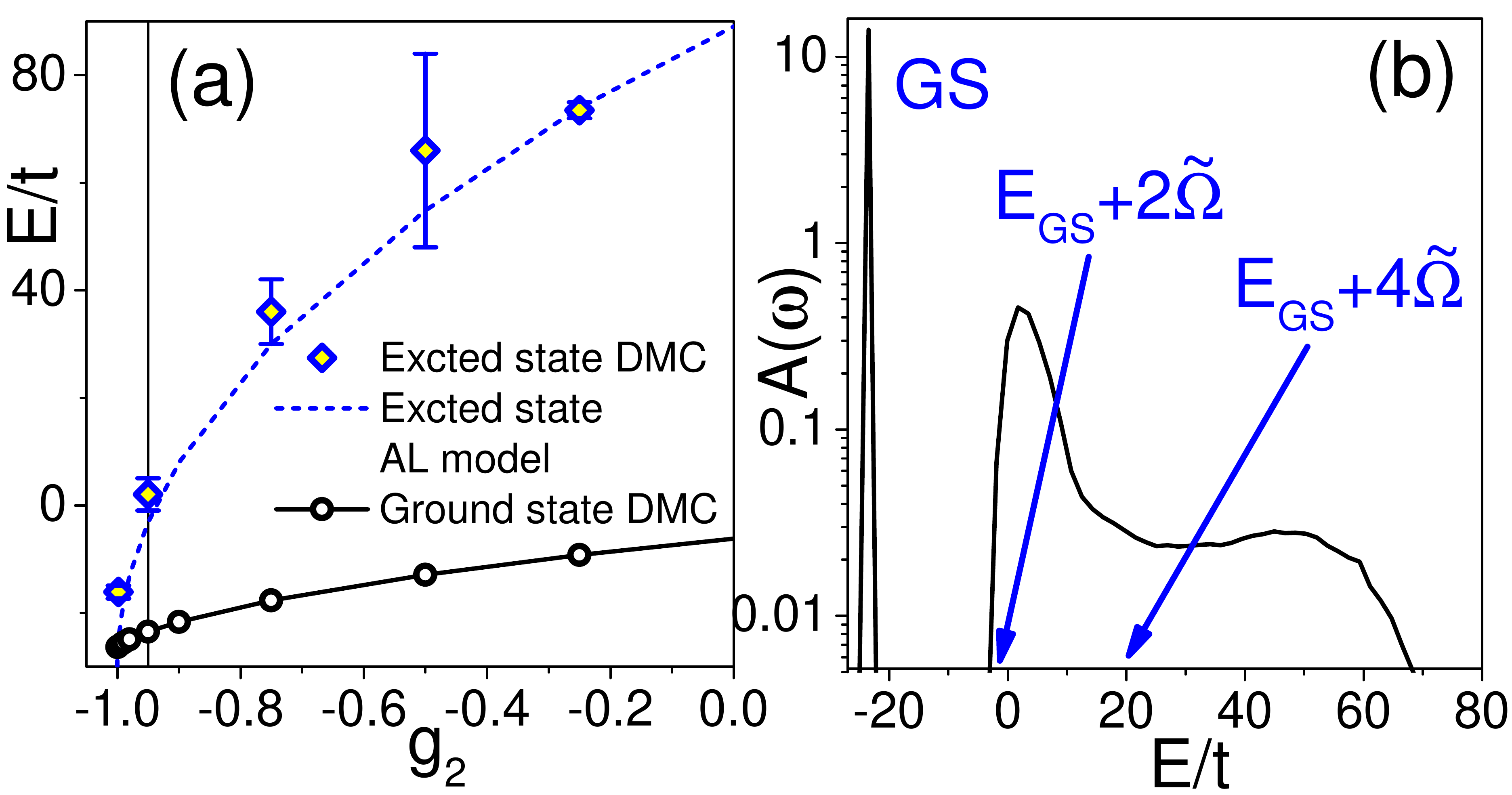}
    \end{center}
    \caption{\label{fig:fig6} 
    Anti-adiabatic limit in 3D at $\Omega/W=4$. (a) Ground state energy (circles connected by the black line) and that of the first excited state (blue diamonds). Blue dashed line is the first excited state energy in the AL. Vertical line at $g_2=-0.95$ corresponds to the coupling for which the spectral function $A(\omega)$ in panel (b) is shown. 
    (b) Spectral function $A(\omega)$ at $g_2=-0.95$ for which  Eq.~(\ref{Or}) predicts $\widetilde{\Omega} \approx 10.7$.  
    Arrows show energies of 2-phonon $2\tilde{\Omega}$ and 4-phonon $4\tilde{\Omega}$ thresholds above the ground state energy.
    }
\end{figure}

To complete the picture, we performed analytic continuation of the Green's function spectral density, $A(\omega)$, 
in the anti-adiabatic limit by the Stochastic Optimization with Consistent Constraints method \cite{MPSS,SOCC2017}.
In Figure~\ref{fig:fig5}b, we show the extracted positions of excited states and how they compare with the AL predictions 
(Fig.~\ref{fig:fig5}a). One can see in Fig.~\ref{fig:fig5}b that the onsets of high energy peaks are well described by 
energies $E_{2k}$ in Eq.~(\ref{AAlimit}). 


\underline{\textit{Conclusions.}}
We find that properties of $X^2$-polarons are dramatically different from those based on the intuition gained during a long history of Holstein polaron studies. 
In the adiabatic regime, $X^2$-polarons are nearly indistinguishable from bare particles for any coupling with $|g_2| \sim 1$. 
In the anti-adiabatic regime, particle properties are renormalized more strongly (but saturate to finite values) when approaching the instability threshold at $g_2=-1$, but remain small for positive $g_2$ except in the limit of large coupling. 
Perhaps the most unexpected result is that the lattice deformation around the $X^2$-polaron remains weak even at the threshold. 
This outcome explains the success of recent work on superconductivity in SrTiO$_3$ \cite{Marel,STO_chandra,STO_kiseliov} which treated electrons as bare particles.    

We established that the adiabatic ratio $\Omega /W$ is the key parameter to pay attention to for this problem, while other microscopic details and even the system dimension are less relevant. 
This fact can be used for the development of approximate schemes, such as momentum average \cite{BerciuPRL2006}, dynamical mean field theory \cite{Vukmirovic}, or many body approach \cite{MBP} in the low density limit, that can then be validated against our numerically exact results. 

The soliton-type solutions \cite{Kuklov1989} cannot form for model parameters simulated in this work. 
The minimal requirement is to have $m_*/m \gg 1$ at $\Omega \ll \tilde{\Omega } \le t $, which is not satisfied even for the ($\Omega /t =0.1$, $\tilde{\Omega }=30 \Omega $) parameter set, for which we find $m_*/m \approx 3$. 
Future work should address the soliton problem under the assumption that the crystal is in close proximity to the quantum critical point when $\Omega \to 0$ and the electron contribution to the local vibrational energy is finite when $\tilde{\Omega } \to $ const.


NP, BS, AK, and ZZ acknowledge support from the National Science Foundation under grants DMR-2032136 and DMR-2032077. NN and ASM are supported by JST CREST Grant No. JPMJCR1874, Japan. MH, SK, and JT acknowledge funding by the Research Foundation - Flanders, projects GOH1122N, G061820N, G060820N, and by the University Research Fund (BOF) of the University of Antwerp. CF, SR, TH, and SK acknowledge support from the Austrian Science Fund (FWF) projects I 4506 (FWO-FWF joint project). The computational results presented have been achieved in part using the Vienna Scientific Cluster (VSC).


\end{document}


\title{Supplemental material for ``Polaron with quadratic electron-phonon interaction''}

\author{Stefano Ragni}
\affiliation{Faculty of Physics, Center for Computational Materials Science, University of Vienna,
A-1090 Vienna, Austria}
\author{Thomas Hahn}
\affiliation{Faculty of Physics, Center for Computational Materials Science, University of Vienna,
A-1090 Vienna, Austria}
\author{Zhongjin Zhang}
\affiliation{Department of Physics, University of Massachusetts, Amherst, Massachusetts 01003, USA}
\author{Nikolay Prokof’ev}
\affiliation{Department of Physics, University of Massachusetts, Amherst, Massachusetts 01003, USA}
\author{Anatoly Kuklov}
\affiliation{Department of Physics \& Astronomy, CSI, and the Graduate Center of CUNY, New York 10314, USA}
\author{Serghei Klimin}
\affiliation{TQC, Departement Fysica, Universiteit Antwerpen, Universiteitsplein 1, 2610 Antwerpen, Belgium}
\author{Matthew Houtput}
\affiliation{TQC, Departement Fysica, Universiteit Antwerpen, Universiteitsplein 1, 2610 Antwerpen, Belgium}
\author{Boris Svistunov}
\affiliation{Department of Physics, University of Massachusetts, Amherst, Massachusetts 01003, USA}
\affiliation{Wilczek Quantum Center, School of Physics and Astronomy and T. D. Lee Institute, Shanghai Jiao Tong University, Shanghai 200240, China}
\author{Jacques Tempere}
\affiliation{TQC, Departement Fysica, Universiteit Antwerpen, Universiteitsplein 1, 2610 Antwerpen, Belgium}
\author{Naoto Nagaosa}
\affiliation{RIKEN Center for Emergent Matter Science (CEMS),
2-1 Hirosawa, Wako, Saitama, 351-0198, Japan}
\affiliation{Department of Applied Physics, The University of Tokyo 7-3-1 Hongo, Bunkyo-ku,
Tokyo 113-8656, Japan}
\author{Cesare Franchini}
\affiliation{Faculty of Physics, Center for Computational Materials Science, University of Vienna,
A-1090 Vienna, Austria}
\affiliation{Dipartimento di Fisica e Astronomia, Universit\`a  di Bologna, 40127 Bologna, Italy}

\author{Andrey S. Mishchenko}
\affiliation{RIKEN Center for Emergent Matter Science (CEMS),
2-1 Hirosawa, Wako, Saitama, 351-0198, Japan}

\maketitle

 \section{Sensitivity to the microscopic details of the electron spectrum}
 
\begin{figure}[h]
    \begin{center}
        \includegraphics[width=8.5cm]{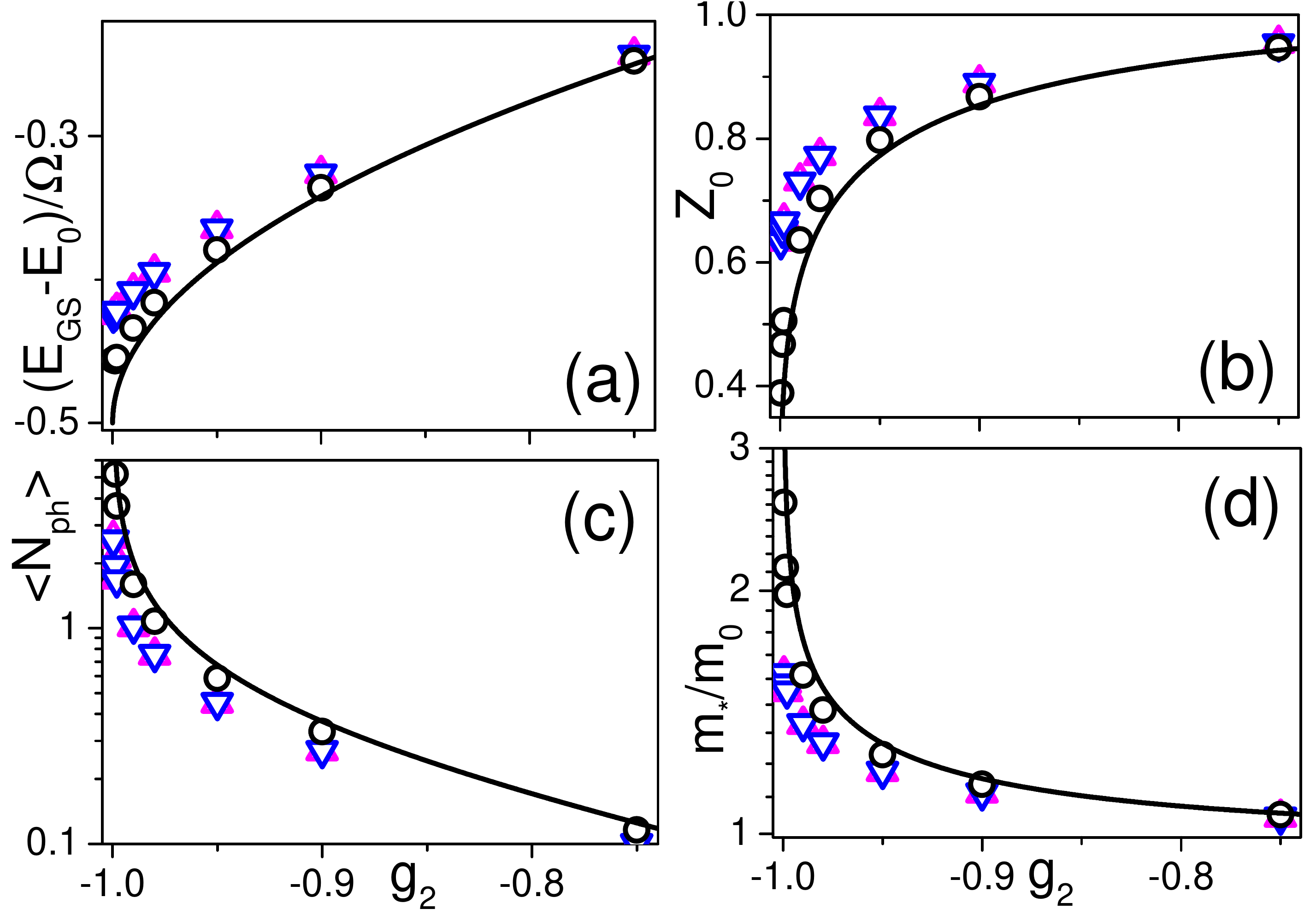}
    \end{center}
    \caption{\label{fig:DimIns}  (a) Energy shifts, (b) $Z_0$-factors, (c) average numbers of phonons $\langle N_{\rm ph} \rangle$, and (d) effective masses $m_{*}/m_0$ for models 1 (triangles up), 2 (triangles down), and 3 (circles). Solid lines show the analytic results in the AL. 
}
\end{figure}

 Figure~\ref{fig:DimIns} demonstrates that the renormalization of $X^2$-polaron properties depends only on the electronic bandwidth and is rather insensitive to the details of the electron dispersion and dimension of space. We consider 
 the anti-adiabatic regime for which all properties change most dramatically with coupling.
 Model 1 is the three dimensional simple cubic lattice with $\Omega/W=4$ and Model 2 is the 
 one dimensional chain with the same adiabatic ratio.   
 Model 3 is the one dimensional chain with the bandwidth reduced by a factor of three, $\Omega/W=12$. 
 Since the three models have different bandwidths $W$ we compare the ground state energy shifts 
 $(E_{GS}-E_0)/\Omega \equiv (E_{GS}-W/2)/\Omega$ in units of $\Omega$.
 
 One can see that ground state energies, $Z$-factors,  average numbers of phonons in the phonon cloud $\langle N_{\rm ph} \rangle $, and effective masses in Models 1 and 2 are nearly impossible to distinguish whereas the results for 
 the Model 3 are different and, as expected, closer to the AL predictions.  
 
 \section{Implementation of the $x$-representation approach}

Here we work in the representation of the eigenstates of local oscillator coordinates $X_i$ (the so-called $x$-representation). A convenient ``visual" starting point is the coordinate path-integral formulation for both the particle and the (local) oscillators. The particle path starts at $\tau=0$ and ends at a certain finite $\tau$. It is used to sample
the ground-state imaginary-time polaron Green's function 
	\begin{equation}
		G(\tau,r) = \langle  a_r(\tau) a_0^\dagger(0)\rangle \, .
	\end{equation}
The next step is a massive integration over the oscillator paths leaving behind only (i) the finite-time propagators in the coordinate representation, (ii) the oscillator ground-state wavefunction, and---only in the case of dispersive phonons---(iii) interaction vertices  between (otherwise local) oscillator coordinates.

In the case of dispersionless phonons---like the one we deal with in the present work, a dramatic simplification comes with the observation that the paths for phonon modes on different sites are disconnected. Hence, we need to sample oscillators paths only for the sites reached by the particle's worldline and need not be sampled 
for time periods when the particle occupation number remains constant.

Without loss of generality, one can count  oscillator energies from their ground state and write
\begin{equation}
H_i^{\rm (ph)} = \Omega b_i^\dagger b_i \, .
\end{equation}
Now one may assume that oscillator paths start at $\tau=-\infty$ and end at $\tau=+\infty$. This way the ``boundary conditions" at $\pm \infty$ become irrelevant:  an infinitely 
long oscillator path is $\tau$-independent and equivalent to the ground-state projector, yielding, upon the integration, the value $U_0(x)$ of the oscillator ground-state wave function at the coordinate $x$ corresponding to the fixed end of the infinite path:
	\begin{equation}
		U_0(x) \, =\, \frac{e^{-x^2/4}}{\left(2\pi\right)^{1/4}} \, .
	\end{equation}

\begin{figure}[h]
		\begin{center}
			\includegraphics[width=8.5cm]{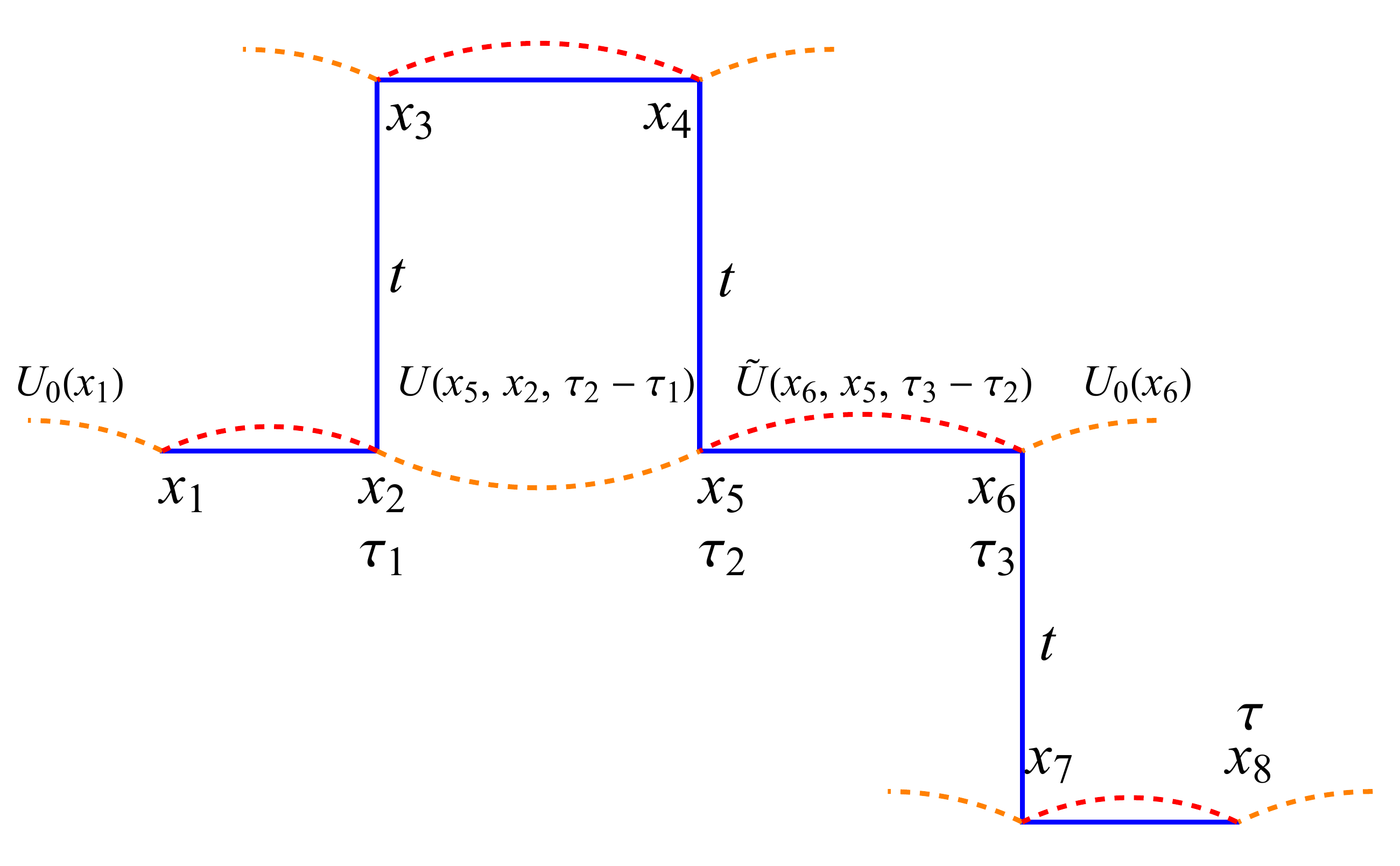}
		\end{center}
		\caption{\label{xreprediagram} A typical diagram for the ground-state Green's function $G(\tau,r)$. Site subscripts are suppressed for clarity. The solid (blue) line represents particle's path, each hopping event of which contributes a factor of $t$. Dashed lines with free ends represent the functions $U_0$. Dashed lines with both ends attached to the particle's path represent $x$-propagators, the choice between $U$ or $\tilde{U}$ is dictated by the absence or presence of the solid line.}
	\end{figure}

Further integration converts the path of the $i$-th oscillator---interacting with the given particle path---into a string of $x$-propagators, $\Pi(x'', x', \tau''-\tau')$, sandwiched between two functions $U_0$ (for clarity, we suppress the site subscript $i$):
\[
U_0(x_1)\Pi(x_2,x_1,\tau_2-\tau_1)\ldots\Pi(x_n,x_{n-1},\tau_n-\tau_{n-1})U_0(x_n).
\]
There are two types of $x$-propagators: the ones corresponding to the free oscillators,
	\begin{equation}
U(x_2,x_1,\tau) \, =\, \langle x_2| e^{-\tau H^{\rm (ph)}} | x_1 \rangle \, ,
 	\end{equation}
and the ones corresponding to the oscillators whose potential energy is changed by the presence of the particle,
	\begin{equation}
\tilde{U}(x_2,x_1,\tau) \, =\, \langle x_2| e^{-\tau \tilde{H}^{\rm (ph)}} | x_1 \rangle \, ,
 	\end{equation}
 	\begin{equation}
 \tilde{H}^{\rm (ph)}\, =\,  H^{\rm (ph)} \, +\,  g_2 {M\Omega^2\over 2} X^2 \, .
 	\end{equation}	
 This way we arrive at the configuration space illustrated in Fig.~\ref{xreprediagram}.
To compensate for an exponential decrease of $G(\tau,r)$ with $\tau$, we use a standard trick of introducing an auxiliary chemical potential $\mu_\tau$.

\begin{figure}[h]
		\begin{center}
			\includegraphics[width=8.5cm]{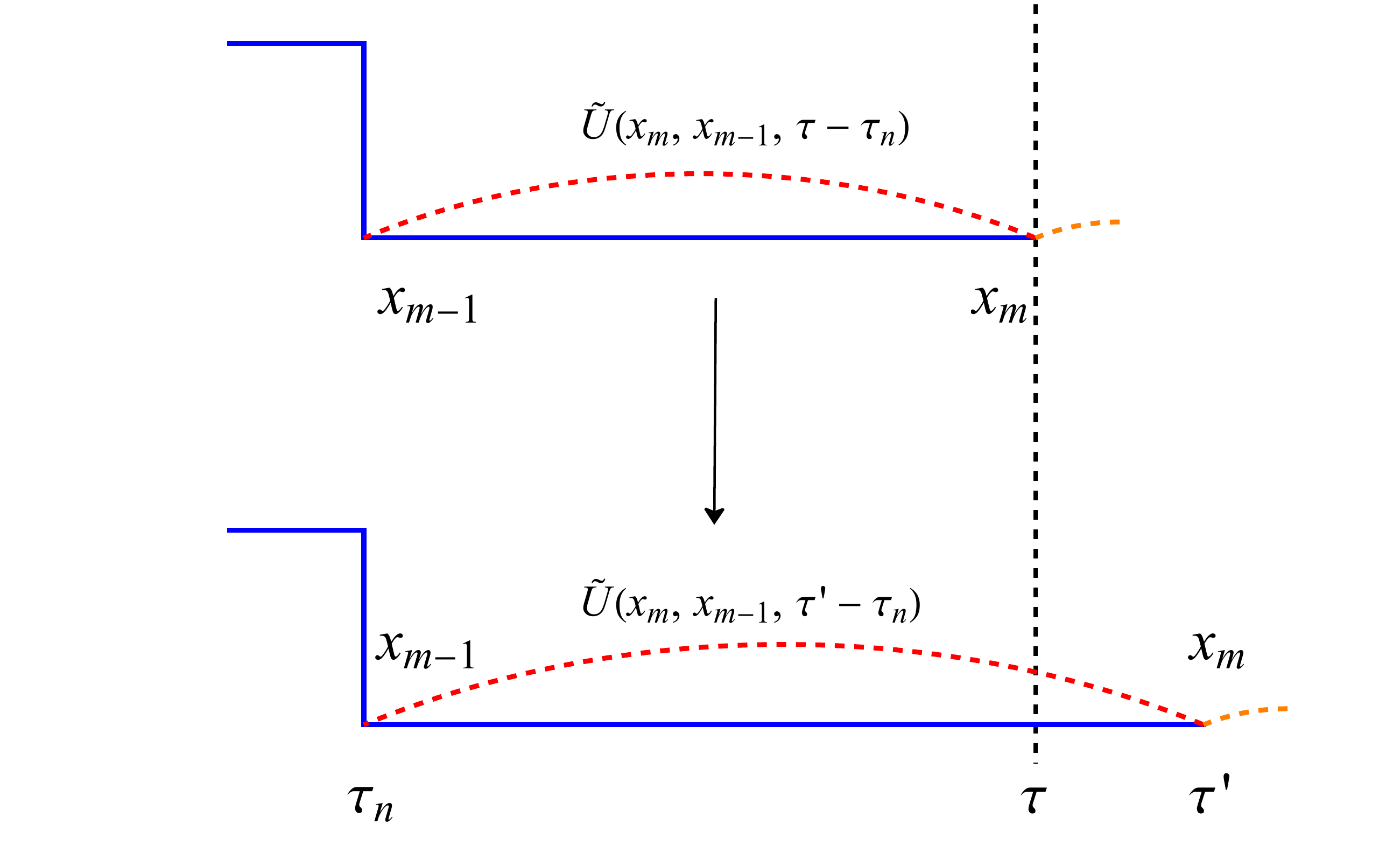}
		\end{center}
		\caption{\label{xrepreUpdateT} An illustrative diagram for the $\tau$-update.}
	\end{figure}

	Our MC scheme includes three updates:
	
	{\bf Change the $\tau$ value}. A new value, $\tau'$, is generated within the interval $[\tau_n, \, \tau_{\rm max}]$ as shown in Fig.~\ref{xrepreUpdateT}., where $\tau_n$ is the  time of the last hopping transition and $\tau_{\rm max}$ 
	is a preset maximum value of $\tau$. The acceptance ratio for this update is given by
	\begin{equation}
			R \, =\,  \frac{P(\tau)}{P(\tau')}\frac{\tilde{U}(x_m,x_{m-1},\tau'-\tau_n)}{\tilde{U}(x_m,x_{m-1},\tau-\tau_n)}e^{-\mu_\tau \left(\tau'-\tau\right)} \, ,
	\end{equation}
	where $P(\tau')$ is the distribution function for generating $\tau'$. 
We choose the exponential distribution $P(\tau') = \mu_p e^{-\mu_p \left(\tau'-\tau_n\right)}$ optimizing the value of $\mu_p$ to make sure that $R \sim 1$.
	
	{\bf Change $x$-value}. A new $x$-value $x_j'$ is generated at a randomly chosen hopping kink. 
	
	Fig.~\ref{xrepreUpdateX} shows two possible cases:
	
	\begin{figure}[h]
		\begin{center}
			\includegraphics[width=8.5cm]{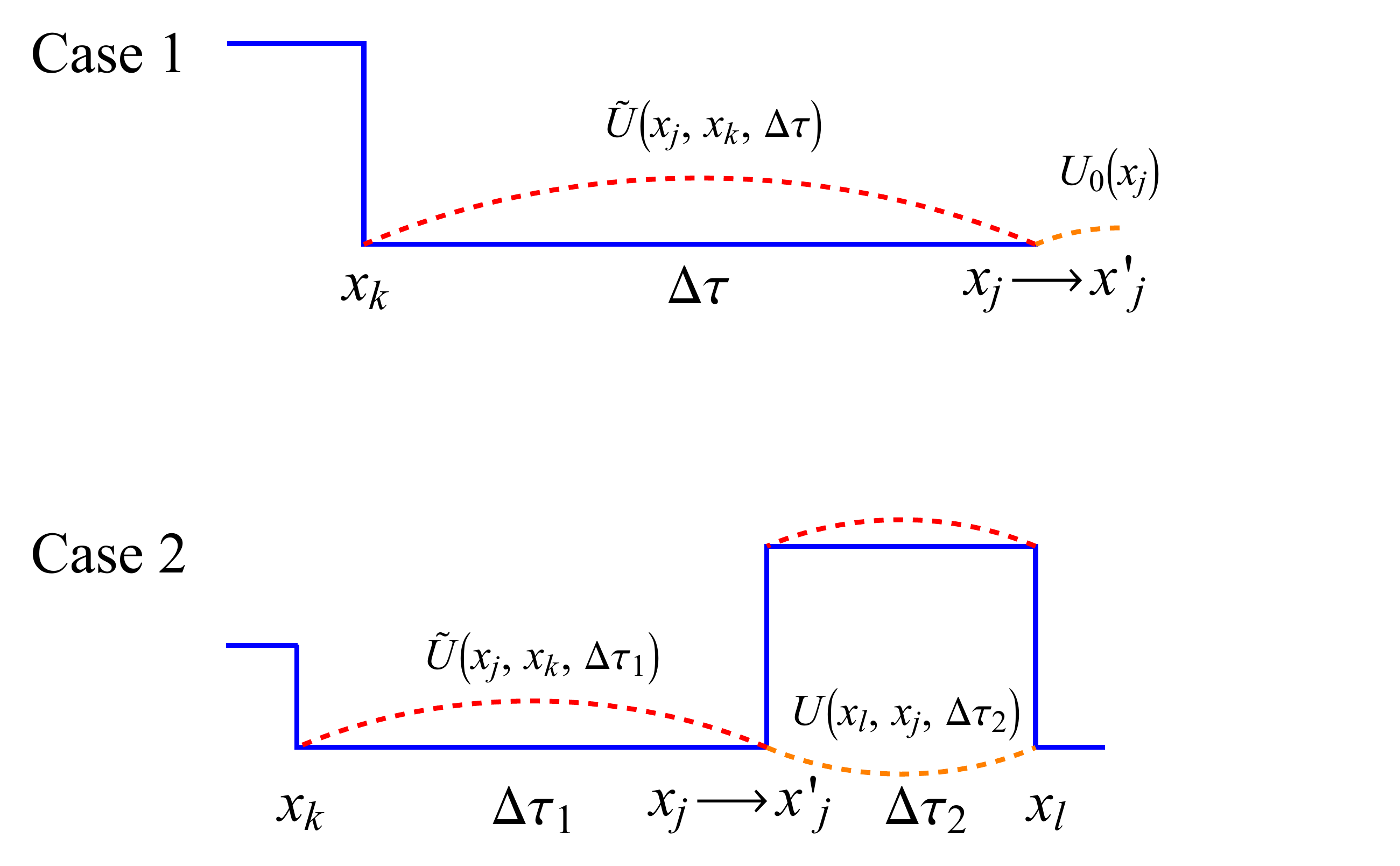}
		\end{center}
		\caption{\label{xrepreUpdateX} An illustrative diagram for the $x$-update.}
	\end{figure}
 
 {\it Case 1.} Here the coordinate $x_j$ is shared between propagators $U_0$ and $\tilde{U}$.  The acceptance ratio is
	\begin{equation}
		R = \frac{P(x_j)}{P(x_j' )}\frac{U_0(x_j')\tilde{U}(x_j',x_k,\Delta \tau)}{U_0(x_j)\tilde{U}(x_j,x_k,\Delta \tau)}
	\end{equation}
	where $P(x_j')$ is the probability distribution function for generating $x_j'$. 
	
	We use the Gaussian distribution
\begin{eqnarray} P(x_j') &=& \left(2\pi\sigma^2\right)^{-1/2} \exp\left[-\left(x_j'-\mu\right)^2/\left(2\sigma^2\right)\right] \, , \label{Gauss} \\
		\mu &=& \frac{r\sigma^2}{2}x_k\, {\rm csch}(\tilde{\Omega}\Delta \tau) \, , \\
	\sigma^2  &=& 2\left[1+r\coth(\tilde{\Omega}\Delta \tau)\right]^{-1} \, ,
\end{eqnarray}
in which case $R = 1$, and the update is always accepted.

 {\it Case 2.} Here the coordinate $x_j$ is shared between propagators $U$ and $\tilde{U}$. The acceptance ratio is
	
	 \begin{equation}
	    R = \frac{P(x_j)}{P(x_j' )}\frac{U(x_l,x_j',\Delta \tau_2 )\tilde{U}(x_j',x_k,\Delta \tau_1)}{U(x_l,x_j,\Delta \tau_2 )\tilde{U}(x_j,x_k,\Delta \tau_1)}\, ,
	 \end{equation}

 	Similarly to Case 1, we achieve $R = 1$ by using the Gaussian distribution (\ref{Gauss}), now  with
 	\begin{equation}
 		\mu  = \frac{\sigma^2}{2}\left[x_l  \, {\rm csch}(\Omega\Delta \tau_2)+r x_k\, {\rm csch}(\tilde{\Omega}\Delta \tau_1)\right]\, ,
 	\end{equation}
 	\begin{equation}
 		\sigma^2  = 2\left[\coth(\Omega\Delta \tau_2)+r\coth(\tilde{\Omega}\Delta \tau_1)\right]^{-1} \, .
 	\end{equation}

\begin{figure}[h]
		\begin{center}
			\includegraphics[width=8.5cm]{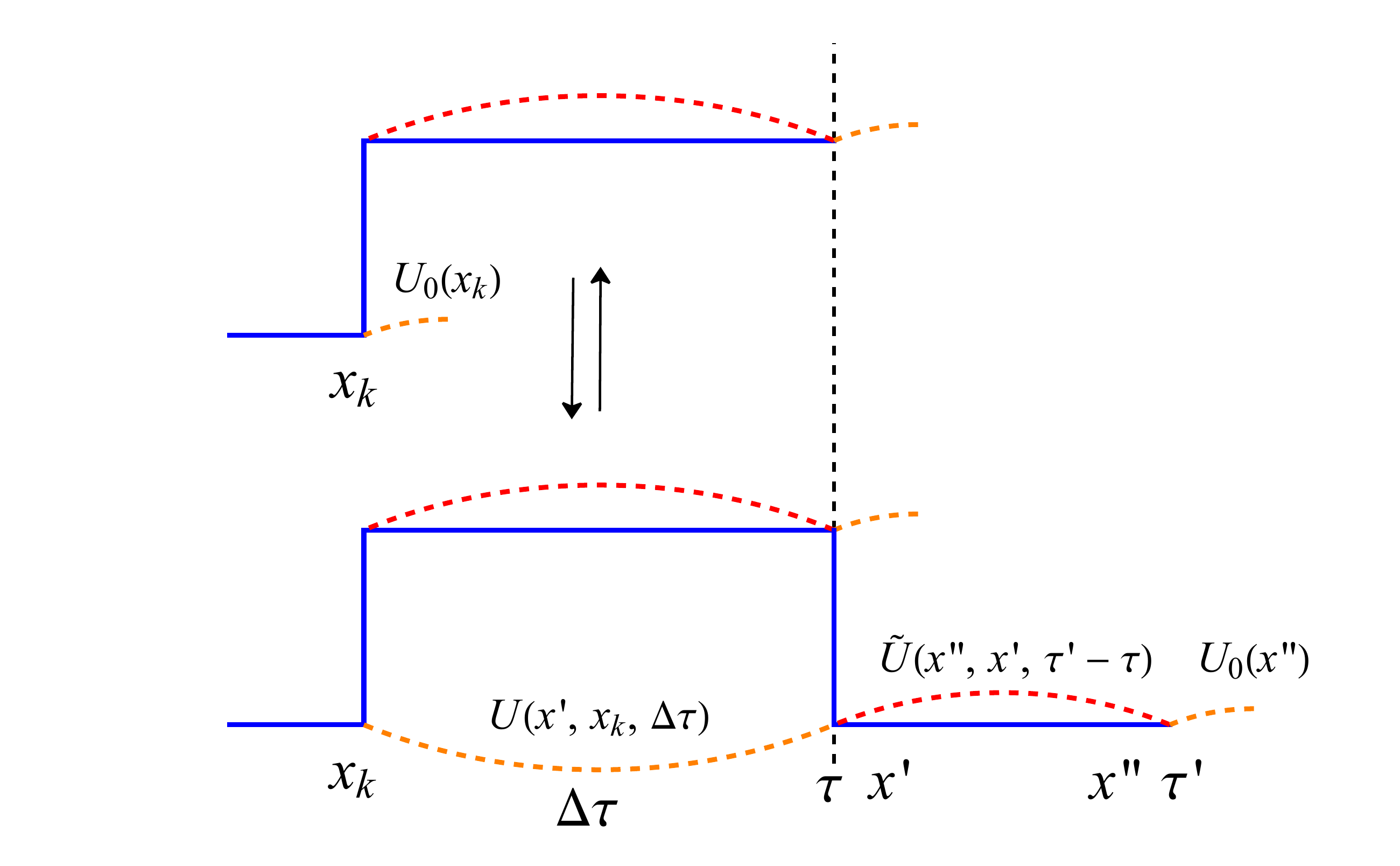}
		\end{center}
		\caption{\label{xrepreUpdateN} An illustrative diagram for adding and deleting hopping kinks.}
	\end{figure}

	{\bf Add/delete a hopping kink}. In this update (illustrated in Fig.~\ref{xrepreUpdateN}), we first  choose to add a hopping kink with probability $P_{\rm add}$ or delete one with $1-P_{\rm add}$. 
	
	{\it Add a kink.} This update adds a hopping transition to any of the $N_{\rm nb}$ nearest-neighbor sites (for 3D cubic lattice, $N_{\rm nb} = 6$) at the end of the  diagram at time $\tau$ and extends the path's imaginary time variable to $\tau'>\tau$. This procedure generates two $x$-values, $x'$ and $x''$, for the  new propagator $\tilde{U}(x'',x', \tau'-\tau)$.
	The acceptance ratio  is
	
	\begin{equation}
		\begin{aligned}
		R_{\rm add} = &\frac{1-P_{\rm add}}{P_{\rm add}}\frac{N_{\rm nb}}{P(\tau')f(x',x'')}te^{-\mu_\tau \left(\tau'-\tau\right)}\\
		&\times\frac{\tilde{U}(x'',x',\tau'-\tau)U(x',x_k,\Delta \tau)U_0(x'')}{U_0(x_k)} \, ,
		\end{aligned}
	\end{equation}
where $P(\tau') $ and $f(x',x'')$ are the distribution functions for generating $\tau'$, $x'$ and $x''$.
	
	There is a special case: if the added hopping transition moves the particle to a site that it never visited before, one should replace $U(x',x_k,\Delta\tau)/U_0(x_k)$ with $U_0(x')$ in the acceptance ratio.
	
	One possible choice for  $P(\tau') $ and $f(x',x'')$ is $P(\tau') = \mu_p e^{-\mu_p \left(\tau'-\tau_n\right)}$ and $f(x',x'')=f(x')f(x'')$ with $f(x) =(2\pi\sigma^2)^{-1/2}\exp\left[-x^2/\left(2\sigma^2\right)\right]$, where $\sigma^2 = 2/(1+r)$.
	
	{\it Delete a kink.} This update deletes the last hopping kink and changes the imaginary time from from $\tau'$ to $\tau$. The acceptance ratio is the reciprocal of the one for the add update:
	\begin{equation}
		R_{\rm delete} = R_{\rm add}^{-1}
	\end{equation}

\section{Implementation of the momentum-space representation approach}

By applying the Fourier transform to the electron and phonon creation/annihilation operators, the Hamiltonian is written in momentum space:
\begin{equation}
\begin{split}
H &= \sum_{\mathbf k} \epsilon_{\mathbf k}a_{\mathbf k}^{\dagger} a_{\mathbf k}^{\,}
+ \Omega \sum_{\mathbf q} b_{\mathbf q}^{\dagger} b_{\mathbf q}^{\,} \\
& + \frac{\Omega}{4} \frac{g_2}{N} \sum_{\mathbf k, \mathbf q_1, \mathbf q_2} a_{\mathbf k + \mathbf q_1 + \mathbf q_2}^{\dagger} a_{\mathbf k}^{\,} (b_{-\mathbf q_1}^{\dagger} + b_{\mathbf q_1}^{\,}) (b_{-\mathbf q_2}^{\dagger} + b_{\mathbf q_2}) \, ,
\label{H}
\end{split}
\end{equation}
where $N$ is the number of unit cells and the lattice constant is set to $a=1$. The electron dispersion is
\begin{equation}
    \epsilon_{\mathbf k} = -2t \sum_{i=1}^{3} \cos(k_i) \, , \qquad -\pi \le k_i < \pi \, .
\end{equation}
We work in the thermodynamic limit, so that momentum sums are turned into integrals:
\begin{equation}
    \sum_{\mathbf k} \to \frac{V}{(2\pi)^3} \int_{\text{BZ}} \! \text{d}\mathbf k \, , \qquad V=Na^3 \, .
\end{equation}

Feynman diagrams in the momentum-space representation are mathematically represented by a diagram weight denoted by the symbol $\mathcal{D}_\text{diag}$. The diagram weight is given by the product of all diagram elements, translated by substituting lines with the electron and phonons propagators presented in Eq.~(3) of the main text. Each interaction vertex is translated to a factor of
$(g_2\Omega/4)/(2\pi)^3$, with the factor $(2\pi)^3$ coming from the thermodynamic limit included for convenience.

\subsection{Combinatorial factors}

In the $S$-matrix expansion, the time-ordered phonon bracket generates many possible Wick pairings. Some of them link the vertices in different ways, so that they are topologically different terms and we represent them as different diagrams. But there are also multiple pairings that are topologically equivalent, and evaluate to the same expression, which means that each diagram may appear more than once in the expansion: that is, it carries a combinatorial factor.

Let us take the simplest case of two vertices at $\tau_A$ and $\tau_B$:
\begin{equation}\label{wick}
    \bra{}\mathcal{T}B_{\mathbf q_1}(\tau_A)B_{\mathbf q_2}(\tau_A)B_{\mathbf q_3}(\tau_B)B_{\mathbf q_4}(\tau_B)\ket{} \, ,
\end{equation}
where $B_{\mathbf q}(\tau) = b^{}_{\mathbf q}(\tau) + b^{\dagger}_{-\mathbf q}(\tau)$. In this case there are two possible Wick pairings (neglecting equal time pairings, which would result in two 1-loops):
\begin{align}
    \contraction{\bra{}\mathcal{T}}{B}{_{\mathbf q_1}(\tau_A)B_{\mathbf q_2}(\tau_A)}{B}
    \contraction[2ex]{\bra{}\mathcal{T}B_{\mathbf q_1}(\tau_A)}{B}{_{\mathbf q_2}(\tau_A)B_{\mathbf q_3}(\tau_B)}{B}
    \bra{}\mathcal{T}B_{\mathbf q_1}(\tau_A)B_{\mathbf q_2}(\tau_A)B_{\mathbf q_3}(\tau_B)B_{\mathbf q_4}(\tau_B)\ket{} \, ,
    \\
    \contraction{\bra{}\mathcal{T}}{B}{_{\mathbf q_1}(\tau_A)B_{\mathbf q_2}(\tau_A)B_{\mathbf q_3}(\tau_B)}{B}
    \contraction[2ex]{\bra{}\mathcal{T}B_{\mathbf q_1}(\tau_A)}{B}{_{\mathbf q_2}(\tau_A)}{B}
    \bra{}\mathcal{T}B_{\mathbf q_1}(\tau_A)B_{\mathbf q_2}(\tau_A)B_{\mathbf q_3}(\tau_B)B_{\mathbf q_4}(\tau_B)\ket{} \, .
\end{align}
Both translate to the 2-loop diagram, which then gets a combinatorial factor of 2.

\begin{figure}[h]
\hfill
\subfigure[above-above and below-below]{
\includegraphics[scale=1]{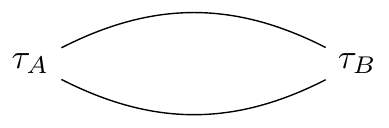}
}
\hfill
\subfigure[above-below and below-above]{
\includegraphics[scale=1]{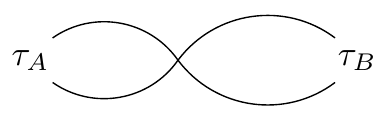}
}
\hfill
\caption{Diagram representation of the two possible Wick pairings for two vertices.}
\label{fig:pairings}
\end{figure}

It is possible to represent these different Wick pairings graphically if we always distinguish between the two ``sockets'' in which we can plug the phonon line, for example above or below the horizontal (see Fig.~\ref{fig:pairings}).

In a general diagram consisting of $N_\text{V} = \sum_{n=2}^\infty n N_n$ quadratic vertices linked together into $N_n$ $n$-phonon loops, the combinatorial factor associated with such diagram is
\begin{equation}\label{fact}
    \boxed{2^{N_\text{V}-N_2}} \; .
\end{equation}
This result can be shown as follows. Consider first the vertices that are linked in 2-loops. Each 2-loop has $2$ possible Wick pairings: (a) and (b) from Fig.~\ref{fig:pairings}. If there are $N_2$ 2-loops, the number of possibilities is $2^{N_2}$.

\begin{figure}[h]
    \centering
    \includegraphics[width=0.48\textwidth]{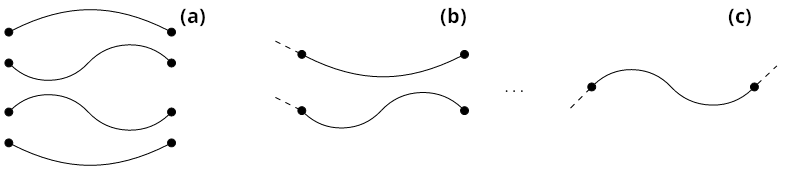}
    \caption{Graphical representation of the factors in Eq.~(\ref{nonpair}).}
    \label{fig:factorproof}
\end{figure}

Consider now a loop of $n$ vertices, with $n > 2$. Start by connecting two of them: this can be done in 4 ways (Fig.~\ref{fig:factorproof}a). Then, for each further vertex added to the loop, there are 2 possibilities (without or with a twist, Fig.~\ref{fig:factorproof}b). The last one is constrained to just $1$ possibility, because it must reattach to the beginning of the $n$-loop (Fig.~\ref{fig:factorproof}c). Putting it all together:
\begin{equation}\label{nonpair}
    4 \times 2^{n-2} \times 1 = 2^{n} \, .
\end{equation}

Multiplying together the contributions from the $N_2$ 2-loops and from higher order loops:
\begin{equation}
    2^{N_2} \times (2^3)^{N_3} \times (2^4)^{N_4} \dots = 2^{N_\text{V}-N_2} \, .
\end{equation}

The result (Eq.~(\ref{fact})) is in agreement with the rule based on diagram symmetries presented in \cite{Goldberg1985}.

\subsection{Updates}
{\bf Add/remove 2-loop}. The addition of a 2-loop to the diagram is proposed. First, one of the $N_\text{V} + 1$ electron propagators is selected at random, and its initial and final times are labeled as $\tau_l$ and $\tau_r$. The initial time $\tau_1$ of the 2-loop is chosen uniformly between $\tau_l$ and $\tau_r$, and the final time $\tau_2$ is chosen from the exponential distribution $P(\tau_2) \propto e^{-2\Omega \left(\tau_2-\tau_1\right)}$. The phonon momenta $\mathbf q_1$ and $\mathbf q_2$ are chosen inside the 1st BZ from the uniform distribution $U(\mathbf q_1, \mathbf q_2) = (2\pi)^{-6}$.

\begin{figure}[h]
    \centering
    \includegraphics[scale=1]{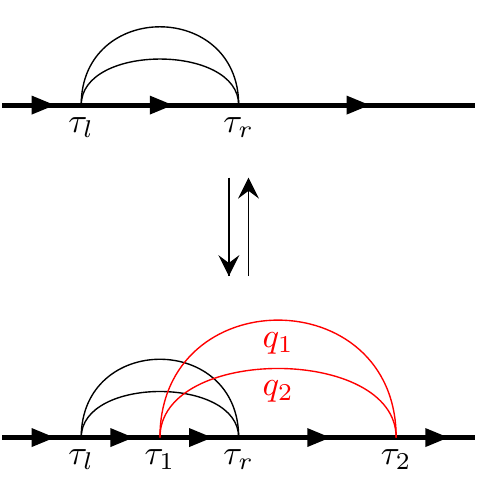}
    \caption{An illustrative diagram for adding and removing 2-phonon loops.}
    \label{fig:add2}
\end{figure}

The update is balanced by proposing the removal of one of the $N_2 + 1$ 2-loops from the diagram, which is chosen uniformly.
The acceptance ratio for the add update is
\begin{equation}
    R_\text{add-2} = \frac{N_\text{V} + 1}{N_2 + 1}\frac{\tau_r-\tau_l}{P(\tau_2)}\frac{1}{U(\mathbf q_1, \mathbf q_2)}\,
    \frac{\mathcal{D}_\text{add-2}}{\mathcal{D}_\text{cur}} \times 2 \, ,
\end{equation}
where $\mathcal{D}_\text{cur}$ and $\mathcal{D}_\text{add-2}$ are the weights of the current and proposed diagram, respectively. The factor of $2$ arises from Eq.~(\ref{fact}) because the proposed configuration contains one 2-loop more than the current configuration. The diagram weight ratio is given by
\begin{equation}
    \frac{\mathcal{D}_\text{add-2}}{\mathcal{D}_\text{cur}} =
    \frac{(g_2\Omega/4)^2}{(2\pi)^6}
    \exp\left\{
    -\Sigma_{\tau_1 \to \tau_2}(\mathbf q_1 + \mathbf q_2)
    \right\} \, ,
\end{equation}
\begin{equation}
    \Sigma_{\tau_a \to \tau_b}(\mathbf q) = \sum_{i \in \tau_a \to \tau_b} [\tilde\epsilon_{\mathbf k_i - \mathbf q} - \tilde\epsilon_{\mathbf k_i} + 2\Omega] \Delta\tau_i \, ,
\label{sigma}
\end{equation}
where the sum is over all electron propagators between time points $\tau_a$ and $\tau_b$, $\mathbf k_i$ is the electron momentum of the $i$-th propagator and $\Delta \tau_i$ its length.

The acceptance ratio of the \textit{Remove 2-loop} update is the reciprocal of the one for the \textit{Add 2-loop} update.

{\bf Add/remove 3-loop}. The addition of a 3-loop to the diagram is proposed. First, one of the $N_\text{V} + 1$ electron propagators is selected at random, and its initial and final times are labeled as $\tau_l$ and $\tau_r$. The initial time $\tau_1$ of the 3-loop is chosen uniformly between $\tau_l$ and $\tau_r$, the middle time $\tau_m$ is chosen from the exponential distribution $P(\tau_m) \propto e^{-2\Omega \left(\tau_m-\tau_1\right)}$ and the final time $\tau_2$ is chosen from the exponential distribution $P(\tau_2) \propto e^{-2\Omega \left(\tau_2-\tau_m\right)}$. The phonon momenta $\mathbf q_1$, $\mathbf q_2$ and $\mathbf q_3$ are chosen inside the 1st BZ from the uniform distribution $U(\mathbf q_1, \mathbf q_2, \mathbf q_3) = (2\pi)^{-9}$.

\begin{figure}[h]
    \centering
    \includegraphics[scale=1]{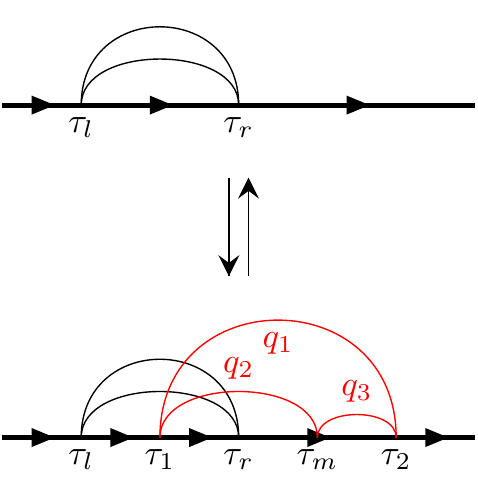}
    \caption{An illustrative diagram for adding and removing 3-phonon loops.}
    \label{fig:add3}
\end{figure}

The update is balanced by proposing the removal of one of the $N_3 + 1$ 3-loops from the diagram, which is chosen uniformly.
The acceptance ratio for the add update is
\begin{equation}
    R_\text{add-3} = \frac{N_\text{V} + 1}{N_3 + 1}\frac{\tau_r-\tau_l}{P(\tau_m)P(\tau_2)}\frac{1}{U(\mathbf q_1, \mathbf q_2, \mathbf q_3)}\,
    \frac{\mathcal{D}_\text{add-3}}{\mathcal{D}_\text{cur}} \times 8 \, ,
\end{equation}
where the factor of $8$ arises from Eq.~(\ref{fact}) because the proposed configuration contains one 3-loop more than the current configuration. The diagram weight ratio is given by
\begin{equation}
    \frac{\mathcal{D}_\text{add-3}}{\mathcal{D}_\text{cur}} =
    \frac{(g_2\Omega/4)^3}{(2\pi)^9}
    \exp\left\{ -(\Sigma_{12} + \Sigma_{13}) \right\} \, ,
\end{equation}
\begin{equation}
    \Sigma_{12} = \Sigma_{\tau_1 \to \tau_m}(\mathbf q_1 + \mathbf q_2) \, ,
    \;
    \Sigma_{13} = \Sigma_{\tau_m \to \tau_2}(\mathbf q_1 + \mathbf q_3) \, ,
\end{equation}
with $\Sigma_{\tau_a \to \tau_b}(\mathbf q)$ defined in Eq.~(\ref{sigma}).

The acceptance ratio of the \textit{Remove 3-loop} update is the reciprocal of the one for the \textit{Add 3-loop} update.

\begin{figure}[h]
    \centering
    \includegraphics[scale=1]{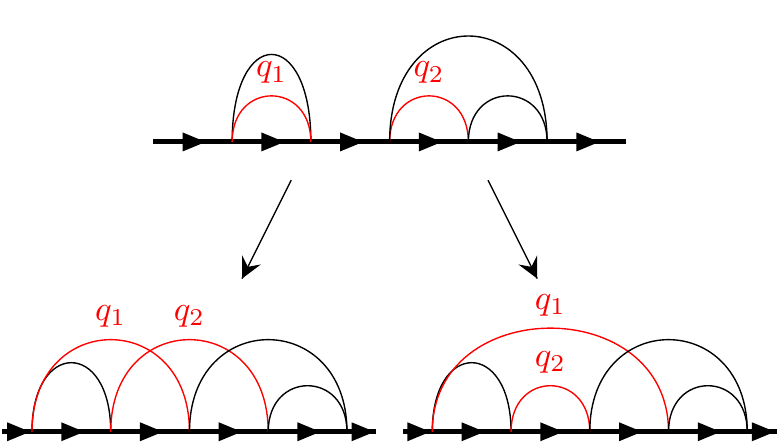}
    \caption{\textit{Relink} update on a diagram containing one 2-loop and one 3-loop. A possible choice for the phonon propagators to relink is highlighted in red, and the possible transitions are shown.}
    \label{fig:relink}
\end{figure}

{\bf Relink}. Randomly pick two phonon propagators across the whole diagram. If the two propagators have any vertex in common, the update is rejected, so that $4$ distinct vertices can be identified. The vertices can be connected in 3 different ways with the phonon propagators leading to 3 distinct diagram topologies (including the topology of the current diagram): we propose at random one of the other two (see Fig.~\ref{fig:relink}).

When evaluating the acceptance ratio, we observe that, unlike in higher order loops, the two phonons in a 2-loop are indistinguishable in the sense that they connect the same vertices. Picking any of them is an equivalent choice for the purpose of the \textit{relink} update, yielding the same final state. For this reason, the probability to switch between configurations with a different number of 2-loops is not balanced.

To satisfy detailed balance, we find the following transition ratio: $2^{\Delta N_2}$, where $\Delta N_2 = N_2' - N_2$ is the difference in the number of 2-loops between the proposed and current diagram.
This ratio exactly compensates the combinatorial factors, so that only the diagram weight ratio is left to compute:

\begin{equation}
    R_\text{rel} = 2^{\Delta N_2} \, \frac{\mathcal{D}_\text{rel}}{\mathcal{D}_\text{cur}} \times \frac{2^{N_\text{V}-N_2'}}{2^{N_\text{V}-N_2}}
    = \frac{\mathcal{D}_\text{rel}}{\mathcal{D}_\text{cur}} \, .
\end{equation}





\subsection{External phonons}

Within the momentum-space approach, information about the phonon cloud can be obtained by extending the simulation to generate diagrams of the irreducible $N$-phonon Green's function \cite{MPSS}. To this end, we generalize the updates previously described in a simple way. When proposing the final time in the \emph{add 2-loop} update, and when proposing the middle and final time in the \emph{add 3-loop} update, we allow them to go beyond the total diagram length $\tau$. The resulting phonon is regarded as an external phonon, as if the representation were circular. Moreover, the \textit{relink} update is generalized to allow phonon propagators to cross the origin in order to create topologies with external phonons.


